\documentclass{JHEP3}
\usepackage{epsfig}
\newcommand{\be}{\begin{equation}}
\newcommand{\ee}{\end{equation}}
\newcommand{\bea}{\begin{eqnarray}}
\newcommand{\eea}{\end{eqnarray}}
\newcommand{\IR}{\mathbb{R}} 

\def\IZ{\relax\ifmmode\hbox{Z\kern-.4em Z}\else{Z\kern-.4em Z}\fi}
\def\IS{\mathbb S} \newcommand{\IT}{{\bf T}}

\newcommand{\non}{\nonumber \\}

\def\half{{1 \over 2}} 

\def\del{{\partial}}
\def\pa{{\partial}}

\def\({\left(} \def\){\right)}
\def\[{\left[} \def\]{\right]}

 \def\co{{\cal O}}
  
\def\cA{{\cal A}}  \def\cC{{\cal C}} 
\def\bC{\bar{\cal C}}

\def\vz{\vec{z}}
\def\tx{\widetilde{X}}
 
 \def\bt{\beta}
  \def\eps{\epsilon}
\def\lam{\lambda}

\def\dmu{\delta\mu}
 
\def\lm{\lambda} \def\lam{\lambda} 
\def\diaglam{\overline{\lambda_i}}
 
\def\tlam{\tilde{\lambda}} 
\def\sgn{\rm{sgn}}
\def\F4GL{F_{4GL}}
 
\newcommand{\sbsection}[1]{\vspace{.5cm} \noindent {\it #1}}

\def\bi{\begin{itemize}} \def\ei{\end{itemize}}

\def\Schw{Schwarzschild }

\def\({\left(} \def\){\right)}
\def\[{\left[} \def\]{\right]}
\preprint{{\tt hep-th/0604015} \\ { \tt NSF-KITP-06-21}}
 
\title{\center{LG (Landau-Ginzburg) in GL (Gregory-Laflamme)}}
 
\author{
Barak  Kol$^1$ and  Evgeny Sorkin$^2$ \\
 $^1$Racah Institute of Physics, Hebrew University, Jerusalem 91904, Israel\\
 $^2$Department of Physics and Astronomy, University of British Columbia\\
 6224 Agricultural Road, Vancouver  V6G 1Z1, Canada \\
{\tt  barak\_kol@phys.huji.ac.il}, ~~~ {\tt  evgeny@physics.ubc.ca }
}
 
\abstract{This paper continues the study of the Gregory-Laflamme
instability of black strings, or more precisely of the order of
the transition, being either first or second order, and the
critical dimension which separates the two cases. First, we
describe a novel method based on the Landau-Ginzburg perspective
for the thermodynamics that somewhat improves the existing
techniques. Second, we generalize the computation from a circle
compactification to an arbitrary torus compactifications.
We explain that the critical dimension cannot be lowered in this
way, and moreover in all cases studied the transition order
depends only on the number of extended dimensions. We discuss the
richer phase structure that appears in the torus case.}
 
\begin{document}

\section{Introduction and summary}
 
In the presence of a compact dimension Gregory and Laflamme (GL)
discovered that uniform black strings are perturbatively unstable
below a certain critical dimensionless mass density \cite{GL1}.
The order of the transition can be computed by following
perturbatively the branch of non-uniform solutions which emanates
from the critical GL string, as first shown by Gubser in the case
of a five-dimensional spacetime \cite{Gubser} where the transition
is first order. That calculation was generalized by one of us (ES)
to arbitrary spacetime dimensions with the surprising result that
the transition is first order only for $D<D^*=``13.5"$ while it is
second order for higher dimensions \cite{SorkinD*}. Here first
order means a transition between two distinct configurations,
while a second order transition is smooth -- the uniform string
changes smoothly into a slightly non-uniform string. Kudoh and
Miyamoto \cite{KudohMiyamoto} repeated the calculation in the
economical Harmark-Obers coordinates \cite{HO1}, confirmed
previous results and observed that in the canonical ensemble the
critical dimension actually changes from $D^*=``13.5"$ to
$D^*_{can}=``12.5"$. All this data constitutes an important piece in the construction of the
phase diagram for this system (see \cite{BKrev} and \cite{HOrev}
for a review).

The present paper includes two main results. First, we show how to
somewhat improve the existing method of calculating the transition
order by employing a Landau-Ginzburg perspective (the basic idea
was described already in Appendix A of \cite{TopChange}).
Secondly, we generalize from the usual $\IS^1 \equiv \IT^1$
compactification to an arbitrary torus compactification $\IT^p$.
 
\sbsection{Landau-Ginzburg improvement to the method}. In Section
\ref{LGsection} we review our set-up and how the description of
phase transitions is achieved by the Landau-Ginzburg (LG) theory,
where one expands the free energy of the system around the
critical point in powers of order parameters. In particular, it is
known that as long as the coefficient of a certain cubic term in
the free energy is non-vanishing then the transition is first
order. If the cubic term vanishes, for instance due to a parity
symmetry such as in our case, then it is the sign of the
coefficient of a certain quartic term, which we denote by $\cC$,
that determines whether the transition is first order or higher
(of course if this term vanishes one has to go to higher terms). A
simpler and more intuitive form of this criterion, which is, alas,
less general and does not apply in our case (as expanded in the
text), is that the transition is second order iff the critical
solution is a (local) minimum of the free energy.
 
Before we can compare the Landau-Ginzburg method with Gubser's
method, we should recall the features of the latter. There one
computes order by order the metric of the static non-uniform
string branch emanating from the critical GL string. The first
order is nothing but the  GL mode. At the second order one
computes the back-reaction. Finally the third order is computed,
or more precisely only the first harmonic along the compact
dimension, from which one can finally compute the leading
coefficients of the changes in mass and entropy,
$\eta_1,~\sigma_2$ of the new branch. The sign of these two
quantities is correlated and determines the order of the
transition.
 
At first sight the two methods look quite different. However, we
show that in the LG method one also needs to precisely compute the
second order back-reaction to the metric. \emph{The third order
however is not required in LG (thereby avoiding the solution of a
set of linear differential equations with sources). Rather one
needs only to expand the action to an appropriate quartic order,
to substitute in the results from the first and second orders and
perform certain integrals that add up to the constant whose sign
determines the order.}

A way to understand the simplification is the following: in
Gubser's method one computes the third order, but it turns out
that all that is really needed is the projection of the third
order onto the GL mode. That is precisely the reason why the first
harmonic sufficed (as the GL mode is in the first harmonic). Our
substitution into the quartic order of the free-energy achieves
exactly that, without the need to compute other properties of the
third order.
 
In Section \ref{sec_T1} we perform the ``Landau-Ginzburg"
calculation for an $\IS^1$ compactification in various dimensions
and verify that we get the same bottom-line coefficients
$\eta_1,~\sigma_2$ as in the previous method, see table
\ref{table_sigma-eta}.
 
\sbsection{Torus compactification}. It is interesting to
generalize the compactifying manifold, and the simplest option
beyond the circle $\IS^1 \equiv \IT^1$, is a product of circles,
or more generally a $p$-torus $\IT^p$. The number of extended
spacetime dimensions is denoted by $d$ and the total spacetime
dimension is $D=d+p$. The critical GL density for such a torus
compactification is easily found to be given in terms of the
shortest vector in the reciprocal lattice \cite{KolSorkin}.
 
We proceed to analyze the transition order in Section
\ref{precalc-section}. First, we motivate restricting ourselves to
square torii. Basically, we view the space of torii as having two
boundaries  -- on the one hand highly asymmetrical torii, where
one (or more) dimensions are much larger than the rest, and on the
other hand highly symmetrical torii such as the square torus.
Since the limit of a highly asymmetrical torus reduces to the case
of a lower dimensional torus (mostly the well-understood case of
$\IS^1$ compactification), we argue that by studying the opposite
limit of a highly symmetrical torus, we achieve an understanding
of both limits and thereby also some understanding of the
intermediate region of general torii.
 
For a square $\IT^p$ torus compactification, $p$ modes turn
marginally tachyonic at the same (GL) point. We find that the
constant $\cC$ is replaced by a $p \times p$ quadratic form
$\cC^{ij}$, in order to allow for the various possible directions
in the (marginally) ``tachyon space'', and that the transition is
second order iff $\cC^{ij}$ is positive for all directions.
Namely, it is enough that there is a single direction in tachyon
space which sees a first order transition for the transition to be
one. Taking into account the $\IT^1$ results we may immediately
deduce that the critical dimension cannot be lower than in the
$\IT^1$ case with the same number of extended dimensions, $d$.
This shows that the indicators of \cite{KolSorkin} for second
order transition at lower dimensions, which were part of the
motivation for this work, were misleading, as further discussed in
the text.
 
Due to the high degree of symmetry of the square torii $\cC^{ij}$
consists only of two independent entries: all the diagonal entries
are the same, denoted by $\cC^=$, and all the off-diagonal entries
are the same, denoted by $\cC^{\neq}$. Since the diagonal term is
precisely the one computed in the $\IT^1$ compactification, we set
to compute the off-diagonal term. Due to the symmetry $\cC^{\neq}$
is the same for all $p$ and for that purpose it suffices to
consider $p=2$, namely we consider the square $\IT^2$ torus. The
only parameter remaining is the number of extended dimensions.
 
In practice we do not compute $\cC^{\neq}$ directly, but rather
the effective $\cC$ for a diagonal direction in tachyon space,
which we denote by $\bC$. Then we solve for $\cC^{\neq}$ which is
a linear combination of $\cC^=,\, \bC|_{p=2}$.
 
In Section \ref{sec_T2} we proceed  to the actual calculation for
$\IT^2$. Once we have chosen the diagonal direction in tachyon
space we are not bothered any longer by the presence of several
(marginally) tachyonic modes. However, the number of metric
components involved in the calculation (back reaction and quartic
coefficient) is larger than in the $\IT^1$ case. Certain discrete
symmetries are found to be helpful in simplifying the calculation.
The calculations and results are described in detail, thereby
applying our improved method to achieve new results.
 
In Section \ref{implications-section} we analyze the results and
their implications. We find that for all the studied values of $d$
where the $\IT^1$ transition is second order, the $\IT^p$
transition is also second order for all $p$. Combining this result
with observation mentioned above we conclude that \emph{the
transition order for square torii shows some robustness in that it
depends only on $d$, the number of extended dimensions, and not on
$p$}.
 
In addition we discuss some subtler implications, including the
finding that for almost all $d$ the diagonal direction in tachyon
space is disfavored relative to turning on a tachyon in a single
compact dimension, and in this sense we have spontaneous symmetry
breaking.
 
The appendices \ref{appendix_eqs}, \ref{appendix_numerics} contain
the details of the full equations, of the numerical technique and summarize various
numerical parameters.
 
\section{Set-up and Landau-Ginzburg theory of phase transitions}
\label{LGsection}

\subsection{Set-up}

The Gregory-Laflamme instability and associated phase transition
physics appear in the presence of compact dimensions, namely for
backgrounds of the form $\IR^{d-1,1} \times Y^p$ where $Y^p$ is
any $p$-dimensional compact Ricci-flat manifold, $d \ge 4$ is the
number of extended spacetime dimensions, and the total spacetime
dimension is $D=d+p$.
 
The simplest compactifying manifold is a single compact dimension
$Y=\IS^1$. We shall first demonstrate the Landau-Ginzburg method
in that context and then turn to $Y=\IT^p$, a square
$p$-dimensional torus. We denote the torus size by $L$ and  work with
Euclidean time of period $\beta$ (corresponding to a canonical
ensemble). Such backgrounds are characterized by a single
dimensionless constant
\be
 \mu_\beta \propto {\beta \over L} ~, \label{def-mu-beta} \ee
where henceforth we shall omit the subscript $\beta$.
 
The non-rotating black objects in which we are interested are
static and spherically symmetric (in the extended dimensions).
After suppressing the time and the angular coordinates the
remaining essential coordinates are $r$, the radial coordinate in
the extended spatial directions, and $z^i, ~1 \le i \le p$ the
periodic coordinates $z^i \sim z^i + L$ which parameterize
$\IT^p$. Thus, the essential geometry has $p+1$ dimensions and we
employ ``cylindrical coordinates'' $(r,z^i)$.
 
\begin{figure}[t!]
\centering \noindent
\includegraphics[width=8cm]{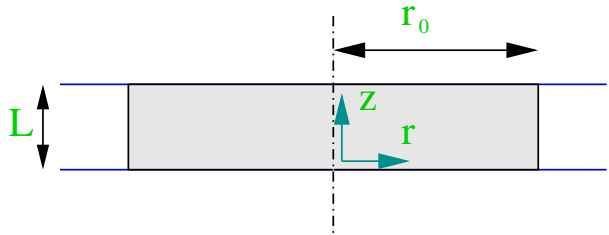}
\caption[]{The uniform black string together with the definition
of the cylindrical coordinates $(r,z)$. $r_0$ is the \Schw
radius.} \label{uniform-string-figure}
\end{figure}

The uniform black $p$-brane\footnote{A 1-brane is often called
``a string''.}, the background around which we perturb (see figure
\ref{uniform-string-figure}) is given by
\be
ds^2 = ds^2_{\mbox{Schw}} + ds^2_Y \label{string-metric} \ee
where $ds^2_Y$ is the metric on $Y^p$, which for our square torus is
\be
ds^2_Y=\sum_{i=1}^p dz^i\, dz^i ~,\label{string-metric2} \ee
and $ds^2_{\mbox{Schw}}$ is the $d$-dimensional \Schw black hole,
 \be
 ds^2_{\mbox{Schw}} =
 -f(r)\,dt^{2}+\frac{1}{f(r)}\,dr^{2}+r^{2}\,d\Omega_{d-2}^{2},
 \label{uniform-string}
 \ee
  where
 \be
 \label{f_schw}
 f(r) = 1-\( \frac{r_0}{r} \)^{d-3} ~,
 \ee %
  and $d\Omega_{d-2}^{2}$ is the metric on the  unit sphere.
 $r_0$ is related to the black hole mass, $M$, via
\cite{MyersPerry} \be
 r_{0}^{d-3} = \frac{16\,\pi\, G_{d}\,M}{(d-2)\,\Omega_{d-2}}
 ~, \label{rho-m} \ee
 where $G_d$ is the $d$-dimensional Newton constant,
and $ \Omega_{d-1}=d { \pi^{d/2} \over (d/2)!} =
\frac{2\,\pi^{\frac{d}{2}}}{\Gamma(\frac{d}{2})}$
 is the area of a unit sphere $S^{d-1}$.
 In Euclidean signature the space-time ends at the horizon.  As
usual, requiring the absence of a conical singularity there fixes
$\beta$, the asymptotic size of the Euclidean time direction, and
it is given by
 \be
 \label{beta_schw}
 \beta = {2\, \pi \over \kappa}={4\, \pi \over f'(r_0)}={4\, \pi\, r_0 \over d-3}~.\ee

Gregory and Laflamme  discovered that the uniform black
string solution (\ref{uniform-string}) develops a $z$-dependent
metric-instability below a certain critical (dimensionless) mass
\cite{GL1}. Equivalently, for a fixed $r_0$
there is a critical wavenumber
\be k \equiv 2 \pi/L \label{wavenumber} \ee
 for instability. At $k=k_{GL}$ the GL mode is marginally tachyonic,
namely a zero-mode.
In \cite{GL1,SorkinD*} the critical GL lengths were obtained for
\Schw black holes in various dimensions (see table
\ref{table_kGL}). From these the high $d$ asymptotic form was
extracted and later proven analytically in \cite{KolSorkin} to be
\be
 k_{GL} \simeq \sqrt{d}\, {1 \over r_0} ~.\label{largeD-GL} \ee
This means that for large $d$ the black string becomes unstable
when $r_0/L \simeq \sqrt{d}$, namely when it is quite ``fat'' and
this indicates that such a string would not decay into a black
hole which would not ``fit'' inside the extra dimension.
\begin{table}[h!]
\centering \noindent
\begin{tabular}{|c||c|c|c|c|c|c|c|c|}\hline
  d& 4& 5 & 6&7&8&9&10&11 \\ \hline
$k_{GL}$  & .876& 1.27& 1.58& 1.85&2.09& 2.30 &2.50 &2.69
\\ \hline
\hline
  d &12&13&14&15&19&29&49&99 \\ \hline
$k_{GL}$  & 2.87 &3.03&3.19 &3.34&3.89 &5.06& 6.72& 9.75
\\ \hline
\end{tabular}
\caption[]{The marginally static
  mode wavenumbers $k_{GL}$ in units of $r_0^{-1}$ as a function of $d$
\cite{SorkinD*,KolSorkin}.}
\label{table_kGL}
\end{table}
 
Having an unstable mode brings us to the issue of the order of the
transition and the existence of additional phases. Rather than
review the state of the art in this respect, we proceed in the
next subsection to describe a somewhat novel method, which in
section \ref{sec_T1} will be shown to reproduce the known results
and then will be used in section \ref{sec_T2} to obtain new
results.

\subsection{Landau-Ginzburg: review and application}
 
The central insight of the Landau-Ginzburg (LG) theory of phase
transitions (see for example \cite{LL_statphys}) is that the nature of a
phase transition can be deduced from the local behavior of the
free energy around the critical solution. This local analysis is
achieved by focusing on the low energy modes and zooming, namely
carrying a Taylor expansion up to a specified order.
For pedagogical reasons\footnote{A brief outline of the method was
already present in appendix A of \cite{TopChange}. } we divide the
discussion into three steps: first we discuss a one-dimensional
configuration space which is the simplest example, then the
general configuration space and finally we utilize the translation
symmetry in $z$ of our background to arrive at the final form of
our formulae.

\sbsection{One-dimensional configuration space}.
The main idea can be demonstrated by a system with a configuration
space consisting of a single variable $\tlam$ and by a control
parameter $\mu$. The thermodynamics is encoded by the free energy
function $F=F(\tlam;\, \mu)$. Generically the free energy does not
have a definite parity in $\tlam$, however sometimes additional
symmetries of the system make the free energy even in $\tlam$,
namely $F(-\tlam;\, \mu)=F(\tlam;\, \mu)$, which will be seen to
hold in our case. This reflection symmetry implies that $\tlam=0$
is an extremum of the free energy (namely that the linear term in
the Taylor expansion of $F$ in $\tlam$ around $\tlam=0$ vanishes).
 Assuming the existence of a critical
solution with a marginally stable mode means that for some
critical value of $\mu$ denoted by $\mu_c$ the quadratic term
vanishes. Thus $F(\tlam;\, \mu) = F_0(\mu)+\cA\, (\mu-\mu_c)\,
\tlam^2 + \dots$ where $\cA$ is some constant, which we take to be
positive without loss of generality (if $\cA$ were negative we
could redefine $\mu \to -\mu$).
 In order to determine the order of the transition it is sufficient
to expand $F$ further as follows
\be
 F(\tlam;\, \mu) = F_0(\mu)+ \cA \, \dmu \, \tlam^2 + \cC \, \tlam^4 +  \dots
  \label{F1d} \ee
 where we defined
  \be
\dmu \equiv \mu-\mu_c
\ee
and $\cC$ is another constant. Actually the whole function
$F_0(\mu)$, is not required for the LG theory, only its expansion
around $\mu \simeq \mu_c$, but we shall not be concerned with this
expansion.
 
\begin{figure}
\centering \noindent
\includegraphics[width=11cm,type=eps,ext=.eps,read=.eps]{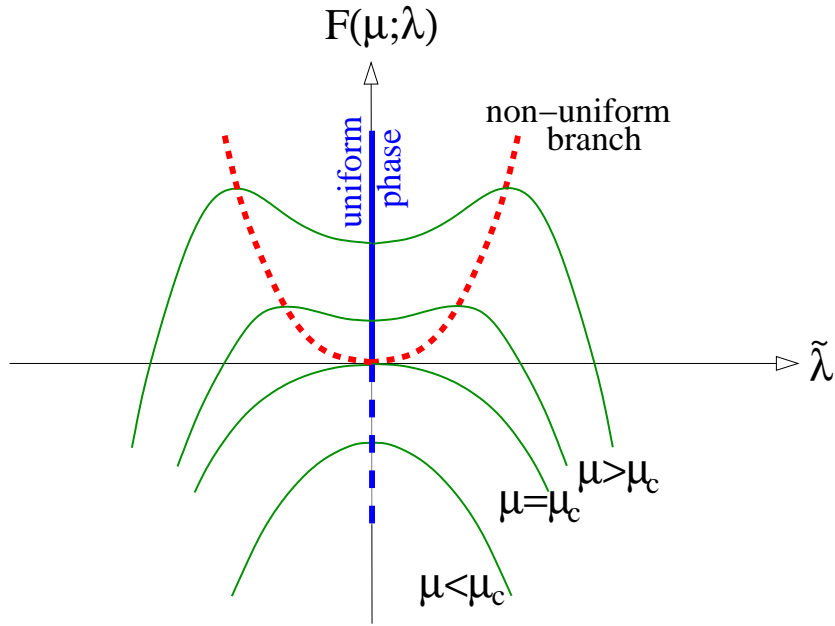}
\caption[]{An illustration of a first order phase transition. A
condensed plot showing two kinds of graphs. The thin solid lines
show the free energy as a function of $\tlam$ for a sequence of
$\mu$ values, while for the thick lines
the vertical axis becomes $\mu$ (the horizontal remains $\tlam$)
and they designate the various phases corresponding to the extremum of the free energy.
The free energy has a minimum for $\mu>\mu_c$
that corresponds to the stable symmetric phase (thick solid line)
which becomes unstable below $\mu_c$ (thick dashed line). It
follows that the asymmetric phase branch emergent from the
critical point (thick dotted line) is unstable since the free
energy has a negative direction for $\mu \ge \mu_c$ at
$\tlam=\tlam_b$.  Note that the free energy in this phase is
higher relative to that of the critical state.} \label{fig_O1F}
\end{figure}

\begin{figure}
\centering \noindent
\includegraphics[type=eps,ext=.eps,read=.eps,width=11cm]{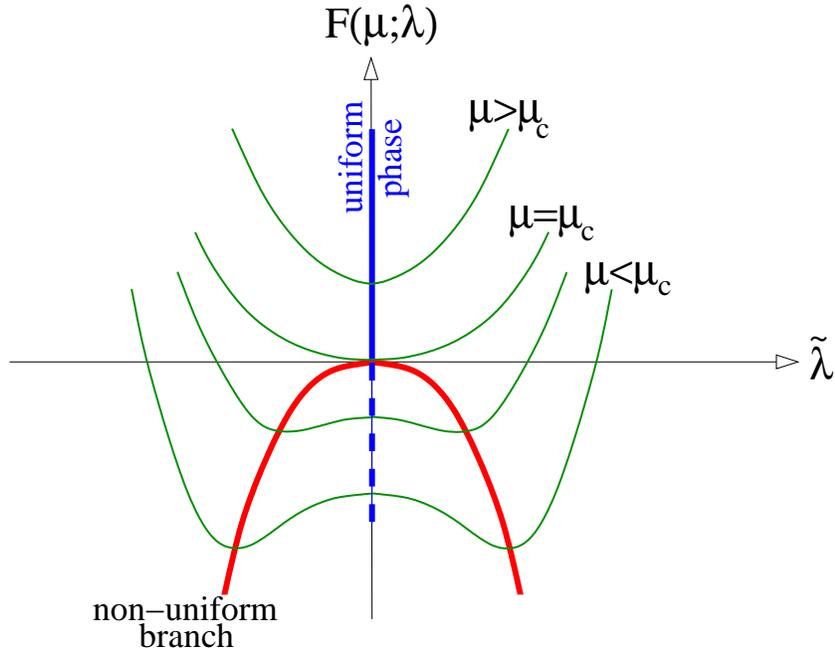}
\caption[]{An illustration of a second order transition from a
symmetric to an asymmetric phase, with the same condensed
conventions as in figure \ref{fig_O1F}.
 The free energy (designated by the
thin solid line) has a  minimum for $\mu>\mu_c$ corresponding to
the stable symmetric phase (thick solid line) which becomes
unstable below $\mu_c$ (thick dashed line). The free energy is at
minimum also for $\mu= \mu_c$ and the minimum continuously moves
away from $\tlam=0$, indicating that the emergent asymmetric phase
is stable and has the asymmetry developing smoothly.}
\label{fig_O2F}
\end{figure}

Once the constants $\cA,\, \cC$ are determined the local
thermodynamics may be deduced by following the form of the free
energy and its extrema as $\mu$ is adjusted in the vicinity of
$\mu_c$. Seeking additional extrema of (\ref{F1d}) through $\del
F/\del\tlam=0$ results in a new branch of extrema in addition to
$\tlam=0$ namely,
\be
\label{lam_b}
 \tlam^2_b= -{\cA \, (\mu-\mu_c) \over 2\, \cC} ~,
 \ee
where $\tlam_b=\tlam_b(\mu)$ describes the new branch.
 Note that since $\tlam^2 \ge 0$ the new branch exists only for
 $\sgn(\mu-\mu_c) = - sgn(\cC)$ (locally). Depending on the sign of
$\cC$ we now get two possibilities which are depicted in figures
\ref{fig_O1F}, \ref{fig_O2F}. For negative $\cC$ the new branch
exists only for $\mu>\mu_c$ and thus cannot serve as the end-state
of the decay for $\mu<\mu_c$ (moreover it is unstable), and
therefore a first order transition will occur into a phase at
finite distance in configuration space. For positive $\cC$ the new
branch can and does serve as the end-point of the decay. The
transition is smooth as one lowers $\mu$ below $\mu_c$. The
difference in free-energies between the two phases is given by
\be
 F_b(\mu)-F_0(\mu)=F(\tlam_b(\mu);\mu)-F_0(\mu)= - {\cA^2\,
\, (\mu-\mu_c)^2 \over 4 \,\cC} = -\cC\, \tlam_b^4 ~. \ee
In particular there is a jump in the second derivative of
$F_b-F_0$ with respect to $\mu$ at $\mu=\mu_c$. Hence by
definition this transition is second order.
 

Given $\cA,\, \cC$ and hence the form of $F=F(\mu)$ , which is the fundamental thermodynamic
potential (in the canonical ensemble) we may find the rest of the
thermodynamic quantities through the usual thermodynamic relations, as we now describe.

We particularize to the black branes in which case the temperature is $T \equiv \hbar/\beta$,
and
so the dimensionless temperature is $\theta=\mu^{-1}$ (\ref{def-mu-beta}).
Rephrasing (\ref{lam_b}) we obtain the leading behavior for the
dimensionless temperature of the emergent branch
\be \label{del_beta} {\delta \theta \over \theta} = - {\delta \mu \over
\mu}=  { 2\, \cC  \over   \cA\, \mu_{GL}  } \tlam_b^2 \equiv \theta_1 \tlam_b^2. \ee

The entropy and the mass of the solutions are
 $S = \mu^2 \del_\mu F$ and $M =\del_\mu(\mu\, F)$.
We can find the coefficients of the second order variations of
these quantities relative to the critical phase
\be
\label{s1}
 s_1 \equiv \( {\cal O}(\lam^2) ~~ {\rm of} ~~  \mu^2 \pa_\mu {F(\mu)-F_0(\mu) \over
 F(\mu)} \)
 = {\mu_{GL} \, \cA  \over d-3},
\ee
 \be
 \label{m1}
 m_1 \equiv \( {\cal O}(\lam^2) ~~ {\rm of} ~~  -\pa_\mu { \mu \[F(\mu)-F_0(\mu)\] \over
 F(\mu)}\)
 = {\mu_{GL} \, \cA  \over d-2}.
 \ee
The variation of the dimensionless mass,
\be
\label{eta}
\eta \equiv G_d M /L^{d-3} =G_d M  \mu^{d-3} /\beta^{d-3}
\ee
in a canonical ensemble for  given $\beta$ is comprised
of two terms arising from variation of $M$ and $\mu$
\bea
  \eta_1 &\equiv& \( {\cal O}(\lam^2) ~~ {\rm of} ~~ {\delta \eta
\over \eta} \) =
 -(d-3) \theta_1 + m_1 = \non
 &=& -2\, (d-3)\, {\cC \over \mu_{GL}\, \cA} + {1 \over
 d-2}\, \mu_{GL} \cA ~.
\label{eta1} \eea

Finally, we can also determine the entropy difference between
non-uniform and the uniform phases with the same mass
\cite{Gubser,SorkinD*}
\bea \label{entropy_dif} {S_{\rm non-uniform} \over S_{\rm uniform}}
&=& 1 + \sigma_1 \lam_b^2 +\sigma_2 \lam_b^4 +\dots, \non
 \sigma_1 =m_1-{d-3 \over d-2} {s_1} &,&
 \sigma_2 =-{d-2 \over 2\,(d-3)^2}\, m_1\, \eta_1  .
\eea
Note that vanishing of $\sigma_1$ is guaranteed by the first law.

\sbsection{Higher dimensional configuration space}.
While the discussion above reveals the essential point, it
considers the free energy to be a function of a single
configuration variable, while in our case it should be considered
a function over the space of all metrics with fixed boundary
conditions, which is an infinite dimensional space. More
explicitly
\be
\label{I_YGH}
 -\beta\, F[g_{\mu\nu}]= I_{GH}[g_{\mu\nu}] \equiv
  {1 \over 16 \pi G_N} \( \int_{\cal M} R + 2 \int_{\del {\cal M}} [K-K^0] \) ~,
\ee
where the first integral is the bulk contribution of the Ricci
scalar and the second is a boundary contribution where $K$ is the
trace of the second fundamental form on the boundary, and $K_0$ is
the same quantity for a reference geometry. The space of metrics
considered is that of all metrics which asymptote to the reference
geometry, which in our case is flat $\IR^{d-1} \times \IS^1_\beta
\times \IT^p_L$  characterized by dimensions $d$ and $p$ and the asymptotic constants
$\beta,\, L$.
 
We denote by $z^i$ the coordinates along the compact dimensions,
and by $r$ the radial coordinate. The most general metric which is
static and spherically symmetric (in the extended dimensions)
\be \label{general-ansatz}
 ds^2 = e^{2\, A}\, dt^2 + ds^2_{(r,z^i)} + e^{2\, C}
 d\Omega^2_{d-3} ~,\ee
where $A$ and $C$ are functions of $(r,z^i)$, $ds^2_{(r,z^i)}$ is
an arbitrary metric on the $(r,z^i)$ space and since the metrics
are static we might as well work with Euclidean signature.

We denote the metric fields collectively by $x$. We are interested
in a background which is the critical GL string, denoted
 by $x^{(0)}$, and we decompose the metric fields into the
 background, $x^{(0)}$,
 and fluctuations, denoted by\footnote{In general $X$ can be considered to be a vector.
While in our case it is infinite dimensional (since the
fluctuations, $X$, are a set of metric functions, and as such
represent infinitely many modes), more generally, if the
configuration space is finite-dimensional then $X$ would be a
finite dimensional vector $X^i$.} $X$
 \be
 x=x^{(0)}+X ~.
\label{back-fluc}
 \ee

Since we wish to follow the new branch emanating from the zeroth
order solution, $x^{(0)}$, we expand in a perturbative parameter
$\lam$
\be
 X =  \sum_{i=1}^\infty \lam^i  X^{(i)} =
\lam\, X^{(1)} + \lam^2\, X^{(2)}+\cdots \label{field-expand}
\ee
The first order contribution must be a zero mode, namely the GL
mode $X_{GL}$
\be
 X^{(1)}=X_{GL}~,
 \ee
and it will be sometimes convenient to denote the second order
term by $X_{BR}$, the back-reaction
\be
 X^{(2)}=X_{BR}~.
 \ee

We need to expand the free energy $F$ to fourth order in $\lam$.
To that purpose we first expand $F$ in $X$
\bea
 F(X)= F_0 &+& F_2(X,X) + F_3(X,X,X)+ F_4(X,X,X,X) +\dots \non
      &+& \dmu\, {\del F \over \del \mu} + \dots     ~,
 \label{FexpandX}
 \eea
where $F_i$ are multi-linear expressions\footnote{More
explicitly, $F_i(X,\dots,X)=\int dr dz F_i(r,z)\, X(r,z) \cdot
\dots \cdot X(r,z)$ where here each $X$ could also be carrying
derivatives.}. The linear term vanishes since by assumption $X=0$
is a solution.
 
Next, we decompose the fluctuations\footnote{More concretely
$\tlam$ is defined by $\tlam :=<X,X_{GL}>$ the projection of a
general perturbation $X$ onto $X_{GL}$ where $<\cdot,\cdot>$ is
the appropriate inner product in the space of perturbations. As
will become clearer in section \ref{sec_T1} the inner product is
given by the coefficient of $k^2$ in the action, and makes the
eigenvalue equation for the GL mode (\ref{a1eq}) self-adjoint. In
the particular  gauge used in this paper it is given by
$<X,\hat{X}>:= const\, \int a\, \hat{a}\, r^{d-2}\, dr$
(\ref{F2T1}). The definition of $\tx$ may be further described as
follows. The spectrum of (\ref{a1eq}) has a single negative mode
$-k_{GL}^2$ and a continuum of scattering states $X_\kappa(r)$
labelled by $\kappa:=-k^2\ge 0$. We may decompose the general
perturbation $X(r)$ according to the orthonormal basis of
eigen-functions as follows $X(r)=\tlam\, X_{GL}(r) + \int_0^\infty
d\kappa\, \tx(\kappa)\, X_\kappa(r)$ where $\tx(\kappa)$ are the
coefficients in this basis.}
 $X$ into the marginal GL mode
$\tlam$, and all the rest, $\tx$ \bea
 X &=& (\tlam,\tx) \non
 X &=& \tlam\, X_{GL} + \tx \label{x-decomp}
\eea
 In terms of the perturbations series ({\ref{field-expand}) we
 have
  \bea
 \tlam &=& \lam  \non
 \tx &=& \lam^2\, X_{BR} + \dots ~. \label{tx-pert} \eea
So $\lam$ and $\tlam$ are equal along the perturbation path, and
the separate notation is intended to distinguish between the
direction in the space of metric fluctuations, $\tlam$, and the
perturbation parameter, $\lam$.
 
We now substitute the decomposition of $X$ (\ref{x-decomp}) into
the $X$-expansion of $F$ (\ref{FexpandX}) keeping only terms
which will end-up being up to fourth order in $\lam$. Noting that
$\tx$ receives its first contribution at the second
order\footnote{$\dmu$ will turn out to be ${\cal O}(\lam^2)$ as
well, see (\ref{lam_b}).}, and taking account of the parity
symmetry for $\tlam$ (which will be justified for our case
shortly) we find
\be
 \rm{\fbox{$~~~
 F(\tlam,\tx;\mu)= F_0(\mu) + \cA\, \dmu\, \tlam^2 + F_2(\tx,\tx) + \tlam^2\,
 G(\tx) + \F4GL\, \tlam^4 ~~~$}}~.\label{F-expand}
  \ee
This is the expression with which the LG analysis is carried out,
and we now proceed to describe its various ingredients. $\cA$ is a
constant that can be read from $\del F/\del \mu$. $F_2$ is the
same bi-linear functional as in
(\ref{FexpandX}),\footnote{Actually, since $X_{GL}$ is a zero mode
 $F_2(X,X)=F_2(\tx,\tx)$ for all $X$.}
 $\F4GL$ is a constant given by
 \bea
\F4GL &:=& F_4(X_{GL},X_{GL},X_{GL},X_{GL})= \non
    &=& \(\co(\lam^4) ~\rm{ of }~ F(\lam\, X_{GL}) \) ~,
    \label{defF4GL}
 \eea
 where by ${\cal O}(\lambda^4)$ we mean the coefficient of ${\cal O}(\lambda^4)$.
 Finally, $G$ is a linear functional given by\footnote{$G(\tx)$ can be described more
  concretely in terms of the source for the back-reaction
 equations $Src_x$ through $G(\tx)=\int dr\, Src_x(r)\, \tx(r)$.}
 \bea
G(\tx)  &:=& 3\, F_3(X_{GL},X_{GL},\tx)= \non
    &=& \(\co(\lam^4) ~\rm{ of }~ F_3(X,X,X)|_{X=\lam\, X_{GL} + \lam^2\, \tx} \) ~.
 \eea

Substituting the perturbations series ({\ref{field-expand}), or
more explicitly (\ref{tx-pert}),
 into the expansion of the free energy
(\ref{F-expand}) we may solve for the back-reaction $X_{BR}$ by
varying the free energy. Equivalently, in practice we expand the
Einstein equations to second order and obtain a source quadratic
in $X_{GL}$. The equations of motion for the back-reaction, may be
symbolically written\footnote{In vector notation it would read $0=2\,F_{2,ij} X_{BR}^j
+  G_i$.}
 as $0=\delta F/\delta X=2\,F_2 X_{BR} + G$
 and its solution\footnote{These back-reaction equations
 are solvable, namely $F_2(\tx,\tx)$ is invertible, since
 the zero mode $X_{GL}$ was removed from its domain of
 definition.} can be written symbolically as $X_{BR}=-(1/2)
 F_2^{~-1} G$.
 
Substituting back the fields
into the free energy (\ref{F-expand}) one obtains the quartic
coefficient $\cC$ as defined above in (\ref{F1d}) to be
\be
 \rm{\fbox{$~~~
  \cC=\F4GL- F_2(BR)    ~~~$}} ~, \label{c-form}
  \ee
where $\F4GL$ was defined in (\ref{defF4GL}) and $F_2(BR)$ is the
quadratic action for the back-reaction defined by
 \bea
 F_2(BR) &:=& F_2(X_{BR},X_{BR}) = \non
          &=& \( \co(\lam^4) ~\rm{ of }~ F(\lam^2\,  X_{BR}) \)
          ~.\eea
Note that the negative sign in front of $F_2(BR)$ comes from the
equation $G = -2\, F_2\, X_{BR}$.

Here we would like to note another more intuitive but less general
interpretation of the criterion for transition order
(\ref{lam_b},\ref{c-form}). In cases where the original phase,
before criticality, was stable, namely it was a local minimum of
the free energy, the criterion for a second order transition is
that the critical solution itself is also a minimum of the action
(despite the presence of the zero mode), since clearly if it is
not a minimum then another lower minimum already exists for a
first order transition to occur. In such a case we may view the
equation for the back-reaction as an attempt to find the direction
of ``steepest decent'' and $\cC$ would be the $\co(\lam^4)$
coefficient of that decent. However, since we work in  the
canonical ensemble for which the (fat) uniform string has a
negative Gross-Perry-Yaffe (GPY) mode \cite{GPY}, the
gravitational action is unbounded from below and the
considerations above do not apply (or apply ``modulo this negative
mode'').
 
\sbsection{Incorporating invariance under $z$-translations}. The
black brane is invariant under the $U(1)^p$ isometry group,
originating from torus translations. As a result we can Fourier
decompose all fields $X$. We can account for this decomposition in
the computation of $\cC$ thereby making (\ref{c-form}) more
explicit. Here we discuss the case $p=1$ (higher $p$ will be
discussed in section \ref{precalc-section}).
 
The $U(1)$ isometry allows us to perform a Fourier decomposition
of all fields as follows
\be \label{Fourier}
 X(r,z)=\sum_n X_n(r)\, \exp(i\, n\, k_{GL} z) ~,
 \ee
where the index $n$ is the harmonic, and in particular $X_0$ is
the $z$-independent mode, $X_1$ is the GL mode etc.\footnote{In
the actual calculation we shall use a slightly different
normalization, expanding into cosines and sines rather than
exponentials, in order to account for $X$ being real.}

The coefficient of the first order mode, $\lam$, is properly
considered to be complex, as follows
\be \lam\, X^{(1)}=\lam\, e^{i\, k_{GL}\, z}\, X_{GPY} \label{lamGPY}
\ee
 where we used $X_{GL}=\exp(i\, k_{GL}\, z)\, X_{GPY}$
 \cite{Reall}. It is seen that the phase of $\lam$ yields
 $z$-translations. By translation symmetry all possible phases of $\lam$ are
equivalent. A natural way to fix the phase is to require $\lam$ to
be real. Choosing real $\lam$ represents the spontaneous breaking of
$z$-translations by the mode, but $\lam \to -\lam$ is a residual
symmetry corresponding to translation by a half period and this is
the origin of the parity symmetry for $\tlam$ which is crucial for
the expansion of the free energy as discussed above.

The free energy is real, therefore $U(1)$ invariance implies the
following simplification
 \bea
F_2(\tx,\tx) &\to& F_2(\tx_0,\tx_0) + F_2(\tx_2,\tx_2) + \dots
\non
 \tlam^2\,  G(\tx) &\to& |\tlam|^2\, G(\tx_0)+\[\tlam^2\,
 G(\tx_{-2})+c.c.\] ~,\eea
where the indices of $\tx$ denotes the harmonic, and the first
equation is a result of the orthogonality with respect to $F_2$
between different harmonics. Altogether the resulting free energy
is
\bea
 F(\tlam,\tx;\mu) &=& F_0(\mu) + \cA \, \dmu\, |\tlam|^2 + F_2(\tx_0,\tx_0) + F_2(\tx_2,\tx_2)
+ \non
     &+& |\tlam|^2\, G(\tx_0)+\[\tlam^2\,
 G(\tx_{-2})+c.c.\] +\F4GL\, |\tlam|^4 ~,\label{F-expand2} \eea
where ``c.c.'' stands for ``complex conjugate''. This form adds
detail to (\ref{F-expand}).
 
Since first order GL mode is (by definition) in the first
harmonic, its square which sources the back-reaction has harmonics
0 and 2, and therefore the back-reaction decomposes into 0 and 2
harmonics. Symbolically \be
 BR=(BR_0,BR_2) \label{BR0BR2} ~.\ee
Finally
\be
  \cC = \F4GL- F_2(BR_0)-F_2(BR_2) ~,
  \label{c-form2}
\ee
is a more detailed version of (\ref{c-form}), where we made use of
the orthogonality of different harmonics.
 
\section{A single compact dimension, $\IT^1$}
\label{sec_T1}
 
In this section we consider backgrounds with single compact
dimension, $p=1$ and $D=d+1$. The most general metric in this case
can be written as
\be
\label{dsT1}
ds^2= e^{2A} f(r) dt^2 + e^{2B} f(r)^{-1} dr^2 + 2 K dr dz + e^{2H} dz^2 +e^{2C} r^2
d\Omega_{d-2}^2,
\ee
where the functions $A,B,K,H$ and $C$ depend only on $r$ and $z$.
If they vanish and $f(r)=1-({r_0/r})^{d-3}$ then the uniform black
string solution is reproduced, where  $r_0$ designates the horizon
location. (Below we set $r_0=1$).
 
We are interested in constructing static perturbations about the uniform black string, hence
the
above metric  functions will be considered as small corrections to the background metric.
 
We must eliminate first the unphysical degrees of freedom by
fixing a gauge.  In our case,  there are two degrees of freedom
(related to diffeomorphisms of the $(r,z)$ plane) that can be used
to eliminate two of the five metric functions in (\ref{dsT1}). It
is not clear what is the ``optimal" gauge choice and naturally we
attempt here to simplify the equations as much as possible. We
choose the gauge partially by requiring $K=0$. The motivations and
fixing of the remaining freedom is described shortly.

\subsection{First order -- marginally tachyonic mode}
\label{sec_T1_O1}
 
The small parameter of the perturbation theory, $\lam$, is the
amplitude of the negative mode. To simplify the derivation of the
equations of motions and expansion of the free energy in this
section we use real $\lam$. This together with the $ U(1)$
symmetry along $z$ suggests that the linear order perturbations
are of the form
\be
\label{O1pert}
X^{(1)}(r,z) = \lambda \, x_1(r) \cos(k z) , \ee
for $X=A,B,C$ and $H$ (see also (\ref{lamGPY}) ).  After plugging
these expressions into the Einstein equations, $R_{\mu\nu}=0$, we
obtain a set of ordinary linear differential equations (ODEs) for
$a_1(r),b_1(r), c_1(r)$ and $h_1(r)$. These equations will
determine the first order perturbations and the critical
wavelength  $k_{GL}$.

At this stage we fix the gauge and our objective is to get the
simplest equations (desirably uncoupled and without singular
points between the horizon and infinity.) \footnote{ One however
should not be too overdiligent here: any gauge can be taken at the
linear order but at higher orders of the perturbation theory the
equations might become degenerate, indicating that the gauge is
too restrictive. For instance in our case one could choose $H=0$.
Then a GPY-sort equation \cite{GPY,Prestidge,KolSorkin} is
reproduced at the linear order; it is easily solved in various
dimensions, yielding $k_{GL}$. Continuing with this gauge to
higher orders of the perturbations theory leads to a contradiction
as the back-reaction equations appear to put constraints on the
first order perturbations. Physically, this is not surprising at
all. Taking constant $H$ is simply inconsistent with allowing
non-uniform solutions, since when the non-uniformity develops  the
scalar charge defined by $H$ must vary \cite{numericI,HO2}.}

Examining the $R_{rz}=0$ constraint,
\be
\label{RrzO1}
2(d-2)f c_1 +r f' a_1 -\( 2 (d-2)f +r f' \)b_1 +r \( a_1'+(d-2) c_1'\)=0,
\ee
we are led to choose
\be \label{gaugeO1}
 B={ 2(d-2)f C +r f' A \over  2 (d-2)f +r f' }.
\ee
In particular, this is our choice at the linear order. From
(\ref{RrzO1}) we get then $ a_1' =- (d-2)\, c_1'$ and hence \be
\label{ac} c_1=-a_1/(d-2), \ee as an integration constant, which
could in principle arise, vanishes by boundary conditions at
infinity. Consequently,  the linear order equations are
\be
\label{a1eq}
  f\,a_1'' +\frac{ (d-2) \,f + r\,f'}{r}\,a_1'  +
  \[ -k^2 + \frac{2\,( d-1)(d -3) \,f'^2}{{\[ 2\, \(d-2\) \,f +
      r\,f'\] }^2} \]a_1 =0,
\ee
\be
\label{h1eq}
  {1\over r^{d-2}} \(r^{d-2} f h_1' \)' +\frac{2\,f - r\,f'
    }{2\,\(d-2\) \,f + r\,f'}\,k^2\,a_1 = 0.
\ee
Besides, there is the constraint
\be
\label{ConO1}
h_1'+ {2\,f-r\,f' \over  2\, \(d-2\) \,f + r\,f'}\,a_1' - \frac{2\,( d-1)(d -3)
\,f'}{{\[ 2\, \(d-2\) \,f + r\,f'\] }^2} \,a_1 =0.
\ee
We do not solve it but we check that it is indeed satisfied by the solution of
(\ref{a1eq},\ref{h1eq}).
 
The equations are subject to boundary
conditions: regularity at the
horizon, that gives
\be
\label{O1bc_hor}
{a_1' \over a_1}=\frac{k^2}{d-3} -2\,(d-2), ~~~~~
h_1'=\frac{k^2}{d-3}\, a_1 ~~~ \textrm{at}~~ r=r_0, \ee
and regularity at infinity, $r\rightarrow \infty$, which  eliminates any
growing solutions.

We begin by solving (\ref{a1eq}). Because of linearity and homogeneity
of (\ref{a1eq}) the value $a_1(1)$ can be chosen freely. A particular
value of $a_1(1)$ defines the {\it normalization} of the negative mode.
We adopt
\be
\label{norm_a1}
a_1(1)=-(d-2).
\ee
Then (\ref{ac}) gives $c_1(1)=1$, the normalization used in
earlier works \cite{Gubser,SorkinD*,KudohMiyamoto}.

Having fixed $a_1(1)$ we are
left with a one-parameter shooting problem: only for a particular
value of the parameter $k_{GL}$ -- the critical GL wavelength -- an
integration outward the horizon converges.  Adjustment of $k$ and
integration are iterated until $k$ is found with desired accuracy. Our
current implementation reproduces exactly the $k_{GL}$-values cited in
the literature, see table \ref{table_kGL}.
 
Next we solve for $h_1$ by integrating (\ref{h1eq}) from the
horizon outwards. A finite solution that asymptotes to a constant
at large $r$'s exists for any choice of $h_1(1)$. Picking some
$h_1(1)$ we integrate, find the asymptotic value of $h_1$, take
minus this value for $h_1(1)$ and reintegrate the equation to have
$h_1$ vanishing at infinity. Namely, the shooting procedure for
$h_1$ reduces to what we dub ``second shot hits''.  Clearly, the
described procedure succeeds because (\ref{h1eq}) contains only
derivatives of $h_1$ so shifting the solution by a constant is
still a solution of (\ref{h1eq}). This completes the first order
computation.

In closing we wish to stress the advantage of the gauge choice
(\ref{gaugeO1}) by comparing our master equation (\ref{a1eq}) for
the negative mode with those considered in the literature. For
example, in Gubser's gauge \cite{Gubser,SorkinD*}, which
eliminates $b_1$ from (\ref{RrzO1}), two coupled equations for
$a_1$ and $c_1$ must be solved simultaneously in order to find
$k_{c}$ (the equations in an arbitrary dimension are listed in
Appendix of \cite{KolSorkin}). Alternatively, in the GPY approach
where one considers perturbations of the d-dimensional \Schw
solution in the transverse-traceless gauge, (which is exactly
equivalent to setting  $H=0$) one obtains a single master equation
for the negative mode \cite{GPY,Prestidge,JGregoryRoss,KolSorkin}.
The equation, however, has an additional singular point apart from
the horizon and infinity, whose presence complicates  the
shooting\footnote{Basically, doing series expansion and matching
  from both sides about this point in a manner described in
  \cite{JGregoryRoss} allows solving the equation quite effectively.
  We have done this, verifying that the eigen-values $k_{GL}$ found
 applying this method are identical with those in table \ref{table_kGL}.}(see \cite{KolSorkin}
for an analytical solution of this equation in a large $d$ limit.)
Recently, the Harmark-Obers gauge \cite{HO1} was considered in
\cite{KudohMiyamoto}  which obtained a single equation free of
additional singular points. This is similar to what we got here.
 
Having a single regular (aside from the boundaries) equation
(\ref{a1eq}) we may arrive at the canonical form $-d^2\psi/d\rho^2
+ V\, \psi = -k^2\psi$ by changing he coordinate to $\rho$,
related to $r$ by $d\rho/dr =f^{-1/2}$, and redefining $a_1$ to
get rid of the first derivative. The equation mimics a
Schrodinger-like problem of a particle with energy $-k^2$ moving
in the influence of potential $V$.  Figure \ref{fig_pot} depicts
the potential and the wave-function of the negative mode in the
case $d=5$.
\begin{figure}[h!]
\centering
\noindent
\includegraphics[width=8cm,type=eps,ext=.eps,read=.eps]{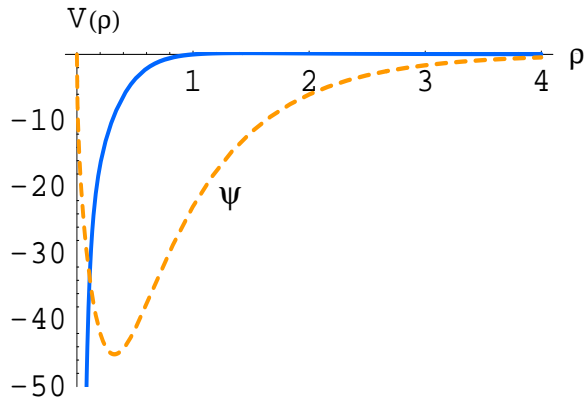}
\caption[]{The potential (solid line) and the wave-function (dashed
  line) for the negative mode, obeying the equation $-d^2\psi/d\rho^2 +
  V\, \psi = -k^2\psi$, for $d=5$.  } \label{fig_pot}
\end{figure}
%
 
\subsection{Back-reaction}
\label{sec_T1_O2}
The back-reaction is comprised of the zero modes and of
the second harmonic modes which we denote as
\be
\label{O2pert}
X^{(2)}(r,z) = \lambda^2 \,  x_0(r)  +  \lambda^2 \,  x_2(r) \cos(2\, k\,  z), \ee
where $X=A,B,C$ and $H$ (see also (\ref{BR0BR2}) ).
 
As usual, in the presence of compact space the modes having
dependence along the internal dimensions decay exponentially with
$r$ and they are massive in the Kaluza-Klein sense. The modes
without $z$-dependence decay as inverse powers of $r$ and they are
massless.
 
At any given order the gauge is given by (\ref{gaugeO1}). However,
the gauge choice in higher orders can be modified by adding terms
from the lower orders with the aim of simplifying the source
terms.

\sbsection{Massless modes}

Our method to simplify the equations is to choose the gauge such
that the constraint $R_{rz}=0$ allows algebraic elimination of one
of the fields.  For the massless modes this constraint is
trivially satisfied since these modes do not depend on $z$
(actually $R_{rz}$ is proportional to $k$ times the LHS of
(\ref{RrzO1}) after replacing the subscripts $1 \to 0$).
 In this case for simplicity we adopt the same gauge as
in (\ref{gaugeO1})
\be
\label{gaugeO20}
b_0={ 2\,(d-2)\,f \,c_0 +r \,f' \,a_0 \over  2 \,(d-2)\,f +r \,f' }.
\ee
Note that the variation of temperature at the
horizon, which is proportional to $\exp(A-B)$, vanishes
automatically in this gauge.

The equations governing the massless mode can be written schematically as
\bea
\label{eqs_T1_O20}
 r^{-(d-2)} \( r^{d-2} f h_0'\)' &=& Src(a_1,a_1',h_1,h_1') \non
{\cal E}_a (a_0'', a_0', c_0',  a_0, c_0)&=&  Src(a_1,a_1',h_1,h_1' ; h_0,h_0') \non
{\cal E}_c (c_0'', a_0', c_0',  a_0, c_0)&=&  Src(a_1,a_1',h_1,h_1' ; h_0,h_0') \non
{\cal Q}( a_0', c_0',  a_0, c_0)&=&  Src(a_1,a_1',h_1,h_1' ; h_0,h_0'),
\eea
where the sources contain squared first order perturbations, the
exact form of the operators $ {\cal E}_a ,{\cal E}_c, {\cal Q}$
and of the sources is found in  Appendix \ref{appendix_eqs}. The
last equation is a constraint -- it is not solved but we verify
that it is satisfied by the solution of other equations in
(\ref{eqs_T1_O20}).
 
The equations (\ref{eqs_T1_O20}) are subject to regularity
boundary conditions at the horizon (\ref{bcO20}).  These
conditions determine the derivatives of the functions in terms of
$a_0(1), c_0(1)$ and $h_0(1)$ which are 3 free parameters. At
infinity we demand the length of the compact circle and the period
of the Euclidean time to reach their unperturbed values, namely we
demand that  $A=H=0$ as $r\rightarrow \infty$.  Besides, an
obvious requirement is that $C$ approaches a constant at large
$r$. However, it turns out that  any choice of that constant
yields regular solution to (\ref{eqs_T1_O20}). In addition, the
constraint  is satisfied for any asymptotic value of $C$ and it
does not contain  any additional information beyond that already
known from the second order equations. Therefore, we end up with
$5$ out of $6$ conditions/parameters which are necessary to
specify a unique solution to the three second order ODEs  in
(\ref{eqs_T1_O20}).
 
The situation is puzzling only until one realizes that the choice
(\ref{gaugeO20}) does not fix the  gauge completely. By analyzing
the residual gauge we find that fixing the residual
reparameterization of $r$ is equivalent to choosing the asymptotic
of $c_0$. To examine the effect  consider a shift
\be
\label{xi}
r\rightarrow r+ \xi(r),
\ee
that induces the variation in the metric functions:
\be
\label{Xxi}
\delta a_0={f'\over 2\,f} \, \xi,~~~~~ \delta c_0= {\xi \over r},~~~~~
\delta b_0 = -{f'\over 2\,f} \,\xi + \xi',~~~~~ \delta h_0 = 0.  \ee
Substituting this into (\ref{gaugeO20}), and demanding invariance
of the gauge  one gets a first order ODE for $\xi$. The equation
has a solution such that $\xi(1)=0$ and that asymptotically $\xi =
\xi_\infty \, r$, with some constant $\xi_\infty$. It follows from
(\ref{Xxi})  that $\xi_\infty$ is the asymptotic value of $c_0$.
Altogether we fix this remaining freedom by requiring $C=0$ at
infinity.\footnote{A similar phenomenon was encountered by Gubser
\cite{Gubser}
 who fixed the residual gauge, remaining after making  the
 conformal ansatz, by assigning $c_0(1)$ some  arbitrary value. Different
choices of this value set different ``schemes" in Gubser's
  language.}

To solve (\ref{eqs_T1_O20}) we first treat $h_0$, which is
decoupled from  other metric functions. The solution is achieved
in the ``second shot hits" fashion since the equation contains
only derivatives of $h_0$, similarly to what occurred at the first
order.  Then we solve two coupled equations for $a_0$ and $c_0$
where $h_0$ as well as squared first order perturbations act as
sources. This is a  two-parameter shooting problem: the values
$c_0(1)$ and $a_0(1)$ are adjusted and the equations are
integrated from the horizon outwards iteratively until a solution
with asymptotically  vanishing $a_0$ and $c_0$ is found.  This
completes the massless-modes computation.

\sbsection{Massive modes}

The equations for these modes are similar to those we had at the
linear order (\ref{a1eq}-\ref{ConO1}) but  this time there are
sources.  To see if we need to modify the  gauge (\ref{gaugeO1})
by adding some lower order terms squared  we examine again the
$R_{rz}=0$ equation:
\bea
\label{RrzO22}
&&2\,(d-2)\,f \,c_2 +r \,f' \,a_2  + 2 \( a_2'+(d-2) c_2'\)+ \non
&&+  {d-1\over 2\,(d-2)} r\,f\, a_1 \,a_1'-\( 2 \,(d-2)\,f +r f' \)b_2  =0.
\eea
From here the obvious choice is
\be
\label{gaugeO22}
b_2={ 2\,(d-2)\,f c_2 +r f' a_2 \over 2 \,(d-2)\,f +r \,f' }
+{d-1\over 2\,(d-2)}{r\,f\, a_1 \,a_1' \over 2\, (d-2)\,f +r\, f' } ,
\ee
since it allows algebraically to eliminate one of the fields, in
precisely the same manner as at the linear order. Consequently,
$c_2=-a_2/(d-2)$ and so we are left with only two second order
ODEs to be solved.
 
Schematically, the equations read
\bea
\label{schem_O22eqs}
{\cal E}_a(a_2'',a_2',a_2)&=&Src(a_1',a_1,h_1',h_1), \non
{\cal E}_h(h_2'',h_2')&=&Src(a_1',a_1,h_1',h_1;a_2)
\eea
and their exact form is listed in (\ref{O22eqs}). These equations
are subject to regularity boundary conditions: both at the
horizon, (\ref{bcO22}), and at infinity.
 
The $a_2$ equation is independent of $h_2$ and it is solved first.
To this end we shoot with $a_2(1)$ as the shooting parameter. It
is adjusted until the regular (decaying at large $r$) solution is
found.  The equation for $h_2$ contains $a_2$ as a source, see
(\ref{schem_O22eqs}) or (\ref{O22eqs}), and once $a_2$ is known
$h_2$ is found by the ``second shot hits'' method. Finally, we
verify that the constraint (\ref{ConO22}) is indeed satisfied by
the solution.
 
\subsection{Free energy and comparison with previous results}
\label{sec_T1_compare}
 
As soon as the negative mode and the resulting back-reaction are found we
are ready to compute the coefficients of the free energy expansion
about the critical point up to the fourth order in $\lam $.
 
The quadratic term in the expansion (\ref{F-expand}), $F_2(X,X)$,
is obtained by expanding the York-Gibbons-Hawking \cite{York,GH}
action integral (\ref{I_YGH}) in our gauge. We find
%
\bea \label{quadI}
  I^{(2)}&=& -{\bt\, \Omega_{d-2} \over 16 \pi G_D}
 \int_0^L dz \int_{r_0}^\infty dr\, r^{d-2}\, {d-1 \over
 (d-2)} \cdot \non
 && \cdot \[ f\, \(\del_r a(r,z) \)^2 + \(\del_z a(r,z) \)^2 -
  {2(d-1)(d-3)\, f'^2 \over \(2(d-2)f + r\, f'\)^2}\, a(r,z)^2 \] ~.
\eea
Note that $h(r,z)$ does not appear here. To get $I^{(2)}$ in terms
of harmonics $a_n(r)$, such as (\ref{O1pert}, \ref{BR0BR2}) one
needs only to substitute in the harmonic decomposition\footnote{Actually, 
we checked this formula only for $n \neq 0$. 
The zero mode sector may contain additional terms.}
$a(r,z)=\sum_{n=0}^{\infty} a_n(r)\, \cos(n\, k\, z)$.

The quadratic part of the action (\ref{quadI}) allows us to
compute $\cA$ (\ref{F-expand}). By definition the integrand
vanishes for $k=k_{GL}$ and $a_1(r)$ being GL mode (found in
subsection \ref{sec_T1_O1}); the leading term appearing as one
moves away from $k_{GL}$, reads
\be \label{I2}
 I^{(2)}= -{\bt\, \Omega_{d-2} \over 16 \pi G_D}
 \int_0^L dz \int_{r_0}^\infty \[2 k_{GL} \delta k {d-1 \over 2\,(d-2)} r^{d-2} a_1(r)^2 \]dr
\lam^2,
 \ee
Using the relations (\ref{def-mu-beta},\ref{beta_schw}) and
(\ref{wavenumber}) we connect the dimensionless wavenumber with
$\mu$ as  $\delta k =2/(d-3) \delta \mu$. It follows then
\be \label{F2T1} \cA \equiv F^{(2)}  /\lam^2 \propto {4\,
\pi\,(d-1) \over (d-2)(d-3) } \int_{1}^\infty a_1^2\, r^{d-2} dr \
. \ee
The factor of proportionality, $ \Omega_{d-2} /( 16 \pi G_D)$,
appears in all action integrals such as (\ref{I2}). So in concrete
calculations we shall get rid of it simply by suitably redefining
the coefficients of the free energy expansion.

The quartic contribution to $F$ is (\ref{c-form2}). We omit the
explicit expressions for the integrand of $\F4GL$ since it is far
more cumbersome then (\ref{quadI},\ref{F2T1}), though it is
straightforward to obtain.
We note that in our gauge (\ref{gaugeO1}) $\F4GL$ includes pieces
which originated in the metric function $B$ before the gauge was fixed.\footnote{It is
also possible to carry the expansion of the action before gauge
fixing, which was the way our computation actually happened to be
performed and only later was it translated into a decomposition of
the gauge-fixed action. The bottom-line result for $\cC$ is of
course the same in both ways.}
 
The numerical values of $\cA$ and of the specific terms forming
$\cC$ are listed in table \ref{table_FT1}. It is evident from this
table that a critical dimension,  $D^*_{can} =12.5$, for the canonical
ensemble appears, above which $ \cC$ changes sign marking a change in
the phase transition's order.
\begin{table}[t!]
\centering
\noindent
\hspace{-1.0cm}
\begin{tabular}{|c|c|c|c|c|c|c|c|c|c|c|c|}
 \hline\hline
  \rm{d} &4&5&6&7&8&9&10&11&12&13&14 \\  \hline
${\mathbf \cA}$   &0.465&2.05&5.11& 9.98& 16.9& 26.2& 37.5& 51.7& 68.7& 88.6& 112\\
${\mathbf \cC}  $ &-0.403& -1.39& -2.99& -5.03& -7.07& -8.47& -8.30& -5.27& 1.97& 15.0& 35.7 \\
$F_{4GL}$ &.0976& .7883& 3.035& 8.417& 19.27& 38.85& 71.42& 122.48& 198.8& 308.8& 462.2\\
$F_2(BR_0)$ &.3918 & 1.287& 2.239& 1.932& -2.112& -13.82& -38.86& -85.06 & -162.4& -283.4& -463.3  \\
$F_2(BR_2)$ &.1076 & .8912& 3.791&11.52& 28.46& 61.14& 118.6& 212.8& 359.2& 577.2& 889.9 \\
\hline\hline
\end{tabular}
\caption[]{Coefficients of the free energy expansions. We notice that $\cC$
  changes sign between $d=11$ and $12$.  This indicates that in the
  canonical ensemble a second order phase transition occurs in dimensions $d \geq 12$. }
\label{table_FT1}
\end{table}

There is a subtlety involved in the calculation of $\cC$.
According to (\ref{c-form}) it is defined as a difference between
two numbers, $F_{4GL} $ and $ F_2(BR)  $.  As the dimension
increases  these numbers grow and they become comparable around
 $D^*_{can}$, see table \ref{table_FT1}.
 Hence, it is imperative to compute each term in (\ref{c-form})
 accurately enough in order to obtain reliable $\cC$.
 
Having computed the coefficients in the free energy expansion we
may obtain the micro-canonical thermodynamic coefficients $\eta_1$
and $\sigma_2$. These are defined in
(\ref{del_beta},\ref{eta1},\ref{entropy_dif}) and their numerical
values are summarized in table \ref{table_sigma-eta}. We include
for comparison the same quantities computed in \cite{SorkinD*}
(see also \cite{KudohMiyamoto}) in Gubser's method. We do not know
which method  is more accurate, but obviously both yield
comparable results (with less than $5\%$ discrepancy.)
\begin{table}[t!]
\centering
\noindent
{\bf Landau-Ginzburg method, derived from $\cA,~ \cC$ } \\
\begin{tabular}{|c|c|c|c|c|c|c|c|c|c|c|c|}
\hline\hline
  \rm{d} &4&5&6&7&8&9&10&11&12&13&14 \\  \hline
$\eta_1  $ &1.40& 3.00& 4.68& 6.21& 7.36& 7.93& 7.68& 6.29& 3.56& -0.695& -6.76 \\
$\sigma_2$ &-0.569& -0.976& -1.40& -1.79& -2.08& -2.21& -2.09& -1.70& -0.969& 0.197& 1.85 \\
\hline\hline
\end{tabular}
 
\vspace{.5cm}
 
{\bf Previous method, direct computation } \\
\begin{tabular}{|c|c|c|c|c|c|c|c|c|c|c|c|}
\hline\hline
  \rm{d} &4&5&6&7&8&9&10&11&12&13&14 \\  \hline
$\eta_1  $ &1.45& 3.03& 4.69& 6.21& 7.34& 7.89& 7.66& 6.25 & 3.57& -0.724 &-6.84 \\
$\sigma_2$ &-0.595& -0.989& -1.42& -1.80& -2.02& -2.14& -2.08& -1.66& -0.966& 0.198& 1.82 \\
\hline\hline
\end{tabular}
\caption[]{Coefficients of the dimensionless mass and of the
entropy variation as defined in (\ref{eta1},\ref{entropy_dif})
computed in both the LG method and the previous method. The match
is within $5\%$.
The sign change (in both quantities) between $d=12$ and $d=13$
indicates a change in the order of the transition in the
micro-canonical ensemble.}
 \label{table_sigma-eta}
\end{table}

Notice that for each dimension both methods produce exactly two
numbers that encode the entire thermodynamics at leading order:
$\eta_1$ and $\sigma_2$ in the micro-canonical ensemble or $\cA$
and $\cC$ in the canonical ensemble, and we have demonstrated the
relation between the two pairs of quantities. We note also that
while in the canonical ensemble $\sigma_1=0$ automatically, in the
micro-canonical ensemble this is a derived result. In fact, in
that case the smallness of $\sigma_1$ can be used as an estimator
of numerical error \cite{Gubser}.

\section{Several compact dimensions}
\label{precalc-section}
 
After having demonstrated the LG-inspired method, we would like to
apply it to the computation of the order of the transition in the
presence of a more general compactification -- by a torus.
 
\sbsection{Square torii}. We choose to restrict our study to
square torii, namely $z^i \sim z^i +L$ with the same $L$ for all
$z^i$ and with right angles between the axes. We motivate this
restriction by the following. We view the space of torii as having
two boundaries  -- on the one hand highly asymmetrical torii,
where one (or more) dimensions are much larger than the rest, and
on the other hand highly symmetrical torii such as the square
torus.
 
For example, for $p=2$ the space of torii is the well known
modular domain parameterized by $\tau$. The highly asymmetrical
2-torus boundary is to be found at $\tau \to i \infty$. The most
symmetrical torii, the square and hexagonal ones are to be found
on the boundary at $\tau=i,\exp(i\, \pi/3)$ respectively. The
boundary of the modular domain also includes other, less
symmetrical torii.
Actually, we could have chosen to study the
hexagonal torus -- it also enjoys a large symmetry that would
simplify the calculations, and our choice is simply one of
convenience.

The limit of highly asymmetrical torii is easy to understand.
Suppose one of the edges of the torus is much longer than the rest
(or equivalently, one of the vectors in the reciprocal torus is
much shorter than others). In this case as the mass of the brane
is reduced the first GL instability to occur will be associated
with this long direction. Since this mode is invariant under
translations in all other directions, and as long as that symmetry
is not spontaneously broken (actually it may very well break to
some extent during the collapse when the highest curvatures
develop) then we find that we can completely ignore all directions
except for the largest one. So effectively that case is reduced to
that of a $\IT^1$ compactification. A similar reduction applies if
there are several directions which are much larger than the rest
-- in such a case we may reduce the problem to one with a smaller
$p$.
 
Therefore by studying the limit of the symmetrical, square torus,
we hope to achieve an understanding of both limits and thereby
also some understanding of the intermediate region of general
torii.
 
\sbsection{GL instability}. In a $\IT^p$ compactification (this
discussion appeared already in \cite{KolSorkin}) we have a GL
instability for each vector in the reciprocal lattice. This
replaces the single instability and its harmonics in the $\IT^1$
case. Therefore as the mass of the black brane is reduced these
instabilities appear in order of smallness of the vectors in the
reciprocal lattice. Of course, once the smallest one is
encountered the system will start collapsing and we will not have
a chance to observe the other instabilities separately.
 
For symmetrical torii, including the square, a non-generic
phenomenon happens: as the first instability is reached (say by
lowering the mass) several modes turn marginally tachyonic
simultaneously. More precisely, for $\IT^p$ there are $p$ such
modes. One may be concerned that this degeneracy occurs only for
exactly square torii, but we would like to argue that this
degeneracy is relevant also for torii which are nearly square.
Indeed, if the torus is not a precise square one of the GL
instability modes will be triggered first. However, as the other
``would be'' tachyons have a very shallow potential at this
moment, once the instability starts ``rolling'' and energy becomes
available there is nothing to stop these modes from getting
spontaneously excited, thereby allowing motion of the system in
all the $p$ ``tachyonic'' directions just as for an exactly square
torus.
 
\sbsection{Generalizing the perturbation method}. We would now
want to generalize the perturbation method and find a
generalization for the coefficient $\cC$ (\ref{c-form}) which
determines the order of the transition. The first order
perturbation is an arbitrary linear combination of the $p$ GL
modes, with coefficients denoted by $\lam_i$, namely the
generalization from $p=1$ to arbitrary $p$ is given by \be
 X^{(1)}=\lam\, X_{GL} \to X^{(1)}=\lam_i\, X_{GL}^i ~,\ee
 where indices $i,j$ are to be summed in the range $1 \le i,j \le
p$ and $X_{GL}^i$ is the GL mode associated with $z^i$ (more
explicitly, the mode computed in subsection \ref{sec_T1_O1} 
after substituting $z \to z^i$).
 
Proceeding to the second order the sources have the following
schematic form $Src \sim \lam_i\, \lam_j\,  X_{GL}^{i}\,
X_{GL}^{j}$. Accordingly the back-reaction becomes \be
 X_{BR}= \lam_i\, \lam_j\, X_{BR}^{ij} ~. \ee
 
To see the implications of the symmetries of the square on
$X_{BR}^{ij}$, let us consider an arbitrary abstract tensor
$T^{ij}$. Since the symmetries of the square allow us to exchange
$z_i \leftrightarrow z_j$ for any $i$ and $j$ it is clear that
there are only two independent components to $T^{ij}$ depending on
whether $i=j$ or $i \neq j$, which we denote as $T^=$ and
$T^{\neq}$ respectively. In summary \be
  T^{ij}= \left\{ \begin{array}{cc} T^= & ~~~i=j \\ T^{\neq} & ~~~i \neq j \end{array} \right.
~.\ee
We shall also make use of another definition, ``T-average''
 \be
 \label{tensor_def}
\bar{T}:= {p\, T^= + p\, (p-1)\, T^{\neq} \over p^2}=
 {T^= + (p-1)\, T^{\neq} \over p}
\ee

Thus there are two kinds of back-reaction. $BR^=$ is the one that
was computed already for\footnote{More precisely, for each
$i$, $BR^{ii}$ is the mode computed in subsection \ref{sec_T1_O2} 
after substituting $z \to z^i$.} $p=1$,  while $BR^{\neq}$ is novel and
will be determined in the next section.
 
Now we must substitute the perturbation series \be
 X= \lam_i X_{GL}^i + \lam_i\, \lam_j\, X_{BR}^{ij} \ee
 into the action and compute the term of $\co(\lam^4)$.
Concentrating on the quartic contribution of the GL modes we find
 the following generalization \be
 \F4GL\, |\lam|^4 \to \F4GL^{ij}\, |\lam_i|^2\, |\lam_j|^2
\ee
 ($U(1)^p$ invariance requires the terms to have equal amounts of $\lam^i$ and
 $\bar{\lam}^i$).
  Similarly, the other factor in the $\cC$ formula (\ref{c-form}),
  $F_2(BR)$ gets generalized as
  \be
 F_2(BR)\,  |\lam|^4 \to
  F_2(BR^{ii},BR^{jj})\, |\lam^i|^2\, |\lam^j|^2 + F_2(BR^{ij},BR^{ij})\, |\lam^i|^2\, |\lam^j|^2 \ee
(note that due to orthogonality of harmonics only the 0-harmonic contributes to
$F_2(BR^{ii},BR^{jj})$ with $i \neq j$).
 
Altogether we find that $\cC$ is replaced by a tensor
\bea
 \cC &\to& \cC(\lam^i)=\cC^{ij}\, |\lam_i|^2\, |\lam_j|^2 \non
 \cC^{ij} &=& \left\{ \begin{array}{cc} \cC^= & ~~~i=j \\ \cC^{\neq} & ~~~i \neq j \end{array}
 \right. ~.
 \label{c-tensor}
 \eea

\sbsection{Determination of the order}. Consider $\cC(\lam^i)$ as
it varies over all possible directions in tachyon space $\lam^i$.
It is enough to consider unit vectors $\sum |\lam_i|^2 =1$,
and thus the vector $|\lam_i|^2$ varies over the simplex given by this
normalization condition together the inequalities $0 \leq
|\lam_i|^2 $ (and $\le 1$). We shall now show that the range of
$\cC(\lam^i)$ is precisely
\bea
 \cC(\lam^i) &\in& [\cC^=,\bC(p)] \subseteq [\cC^=,\cC^{\neq}] \label{c-range} \\
 \bC(p) &:=& {\cC^= +(p-1)\,\cC^{\neq} \over p} \label{def-cbar}
\eea
where $[a,b]$ denotes here the interval between $a$ and $b$
irrespective of which one of $a,\,b$ is bigger.
To see that we compute
\bea
 \cC(\lam^i) &=& \cC^{ij}\, |\lam_i|^2\, |\lam_j|^2 = \non
            &=& \cC^=\,  \sum |\lam_i|^4 + \cC^{\neq}\, \( \(\sum
            |\lam_i|^2\)^2 - \sum |\lam_i|^4\) = \non
            &=& \cC^{\neq} + \( \cC^= - \cC^{\neq} \) \sum
            |\lam_i|^4 ~.\eea
Since $\sum |\lam_i|^4$ ranges precisely over $[1/p,1]$
(\ref{c-range}) follows.
 
The transition is second order if and only if $\cC(\lam^i)$ is
positive for all directions in tachyon space $\lam^i$, for
otherwise, if there exists a vector $\lam^i$ such that
$\cC(\lam^i) <0$ the system will spontaneously settle on that
direction and a first order transition will ensue. We conclude
that \emph{the transition is second order if and only if both}
\be
 \cC^=,\, \bC(p) \ge 0. \label{order-cond}
\ee

\sbsection{$p$-dependence from $p=2$}. From their definition it is
evident that $\cC^=,\, \cC^{\neq}$ are $p$-independent. Actually
$\cC^=$ is already known from $p=1$
\be
 \cC^= = \cC
\ee
Moreover, it suffices to compute $\cC^{\neq}$ for $p=2$ where the
mixed $ij$ terms appear -- the computation does not change for
higher $p$.

 Thus it remains to determine $\cC^{\neq}$. Rather than determine
it directly we take a somewhat more physical approach and pick the
diagonal direction in $\lam$ space, namely
 \be
 \overline{\lam_i} = {1 \over \sqrt{p}}\, \lam \label{diag}
 \ee
independently of $i$. $\cC( \diaglam )$  computed along this
direction is precisely $\bC$ defined in (\ref{def-cbar}).
Therefore by
computing $\bC|_{p=2}$ we obtain
\be
 \cC^{\neq} = 2\, \bC|_{p=2} - \cC^= ~.
 \label{cneq}
 \ee

The $\cA$ term in the free energy may be easily generalized as
follows \be
 \cA\, \dmu\, |\lam|^2 \to \cA\, \dmu\, \lam^i\, \bar{\lam}^i ~,\ee
 and on the diagonal direction (\ref{diag}) $\bar{\cA}=\cA$.
 
The actual computation (at $p=2$), which is fully described in the
next section, requires obtaining the mixed term $\F4GL^{\neq}\,
|\lam_1|^2\, |\lam_2|^2$ from substitution into the action, a new
mixed term in $F_2(BR_0)$, computing the mixed back-reaction term
$BR^{\neq} \equiv BR_{11}$, which is in the $(\pm 1,\pm 1)$
harmonics, and finally substituting it into the quadratic action
to obtain $F_2(BR_{11})$. All the other ingredients were
essentially computed already for the $p=1$ case.
 
\sbsection{Micro-canonical ensemble}. Having obtained the
``diagonal" coefficients $\bar{\cC}$ and $\bar{\cA}$ we can
compute the variation in dimensionless mass $\bar{\eta}$
(\ref{eta}) along this direction by using (\ref{eta1})
\be
 \bar{\eta}_1 := -2\, (d-3)\, {\bC \over \mu_{GL}\, \cA} + {1 \over
 d-2}\, \mu_{GL} \cA
  \label{eta1-bar}
\ee
In fact, we can define a tensor of mass variation $\eta_1$, as in
(\ref{tensor_def}). Then, in analogy with what we had for the
tensor $\cC$, we have $\eta_1^{=} \equiv \eta_{1}|_{p=1}$, and
recalling (\ref{cneq}), $\eta_1^{\neq}= 2\, \bar{\eta}_1|_{p=2}
-\eta_1^{=}$. By analogy with (\ref{order-cond}) the transition
will be second order in the canonical ensemble for a given $p$
exactly when both
\be
 \eta_1^{=},\bar {\eta}_1(p)  <0 ~,
  \label{micro-can-order-cond}
   \ee
where as before (\ref{def-cbar}) we define
$\bar{\eta}_1(p):=\eta_1^= /p+(p-1)\, \eta_1^{\neq}/p$.

\section{Computation for $\IT^2$}
\label{sec_T2}
In this section we perform the computation for the square
two-torus, so $D=d+p$ with $p=2$.
 
The most general metric consistent with the $U(1)_t \times
SO(d-1)_\Omega$ isometries is
\be \label{metricT2} ds^2 =f(r)\, e^{2\,A} dt^2+ f(r)^{-1}\,
e^{2\,B} dr^2 + K_i\,dz^i\,dr + H_i\, dz^i dz^i  +W\, dz^1 dz^2 +
r^2\, e^{2\,C} d\Omega_{d-2}^2 \ee
where $i=1,2$.  The metric in the $(r,\vec{z})$ space has $3$
gauge  degrees of freedom. We fix $2$ of them by choosing $K_i=0$.
The remaining degree of  freedom will be fixed by considerations
similar to the $\IT^1$ case.
 
The square torus has also the following three discrete isometries,
which will also be preserved by our perturbation
\bea
 z_1 &\leftrightarrow& z_2 \non
 z_1 &\to& -z_1 \non
 z_2 &\to& -z_2 \label{symmetries} \eea
%
 
\subsection*{First order -- marginally tachyonic modes}
\label{sec_T2_O1}
 
In the square torus background two GL modes turn marginally
tachyonic simultaneously, resulting in a two-dimensional
``tachyon space". In this section we focus on the diagonal
direction in the ``tachyon space", given schematically by
$\overline{\lam_i}\, X_{GL}^i$ where
\be
 \overline{\lam_i}= \lam
 \label{diag2}
 \ee
Here it is convenient for us to use this normalization to compute
the back-reaction and turn to the normalization (\ref{diag}),
which differs by a $\sqrt{2}$ factor, later when we substitute
into the free energy (which amounts to dividing the quadratic and
the quartic coefficients of the free energy by $p=2$ and by
$p^2=4$, respectively).

As in $\IT^1$ case we use real $\lam$. The diagonal instability is
given explicitly in terms of the functions $a_1(r),\, b_1(r),\,
c_1(r),\, h_1(r)$ and the critical constant $k \equiv k_{GL}$
which were found in subsection \ref{sec_T1_O1} as follows
\be \label{O1Tp_1} X(r,z)=\lm\, x_1(r)\, \sum_{i=1}^2 \cos(k
\,z_i), \ee
for $X=A,B,C$ and
\bea \label{O1Tp_2}
 H_i(r,\vec{z})&=&\lm\, h_1(r) \cos(k\, z_i), \non
 W&=&0. \eea
Clearly, because of the symmetry between the tachyons along the
torus edges only a single function $h_1(r)$ is used for both
$i$'s, and actually the relation $z_1 \leftrightarrow z_2 \Rightarrow H_1
\leftrightarrow H_2$ will hold throughout the perturbation, so
essentially the $H$'s reduce to a single function. Note that the
discrete square symmetries (\ref{symmetries}) are indeed preserved
by the perturbation.

As expected, after plugging the expansions into the Einstein
equations, the equations for different $i$'s completely decouple.
Besides, the gauge conditions and the equations are identical for
every $i$ and they coincide with those derived in the $\IT^1$
case, see  (\ref{a1eq}-\ref{ConO1}). The upshot is that the first
order computation in the  $\IT^2$ ($\IT^p$ in fact) case is
essentially identical with $\IT^1$.
 
\subsection*{Back-reaction}
\label{sec_T2_O2}
 
The form of the back-reaction sources can be roughly obtained by
squaring the linear order expansions (\ref{O1Tp_1}) and
(\ref{O1Tp_2}). In addition, a mixed $W$ term, which was absent in
the $\IT^1$ case, becomes possible.

\vspace{0.5cm}
\sbsection{Massless zero modes}
 
The expansion here is essentially the same as for $\IT^1$.
\bea \label{O20Tp_1}
 X(r,\vz)&=& 2\, \lam^2\, x_0(r) ~,  \non
 H_1=H_2 &=& \lam^2\, h_0(r) ~, \non
 W &=& 0 ~, \eea
where $X=A,B,C$ and the $p=2$ factor comes from summing over the
back-reaction zero-modes for all $i$. The vanishing of $W$ is
dictated by the square symmetries (\ref{symmetries}).
 
To confirm this form we proceed as follows. We use the gauge
(\ref{gaugeO20}). The ODEs obtained after substituting the above
expansion into the Einstein equations are given by
(\ref{Eqh0},\ref{Eqac0}). The equations differ from the $\IT^1$
case only by the appearance of the factor $p=2$ that multiplies
the sources in (\ref{Eqac0}). As a result, $h_0(r)$ is unchanged
while $a_0$ and $c_0$ are modified -- multiplied by $p=2$ relative
to $\IT^1$ solutions.
Indeed, that was the reason that we introduced the factor of $2$
in (\ref{O20Tp_1}).

\sbsection{Massive second harmonic modes}
 
The treatment of these modes replicates the one we had for $p=1$.
The obvious  expansions
\be \label{O22Tp_1} X(r,\vz)=\lam^2\, x_2(r)\, \sum_{i=1}^2\cos(2 k
\,z_i), \ee
for $X=A,B,C$ and \bea \label{O22Tp_2} H_i(r,\vec{z})&=&\lam^2\,
h_2(r) \cos(2 \,k\, z_i), \non W&=&0, \eea
yield exactly the same equations as for the one-torus
(\ref{O22eqs}). So nothing new needs to be solved.
 
\sbsection{Massive mixed modes}
 
This is the longest section here, for these modes are the only new
ones  that were not discussed in the $\IT^1$ section.
 
The only mixed modes expansions allowed by the symmetries
(\ref{symmetries})  are
\be \label{O211Tp_1} X(r,\vz)=\lam^2\, x_{11}(r)\,\cos(k
\,z_1)\cos(k\, z_2), \ee
for $X=A,B,C,H$ and \be \label{O211Tp_2} W(r,\vz)= \lam^2
\,w_{11}(r) \,\sin(k \,z_1) \sin(k \,z_2). \ee
Plugging this into the Einstein equations we examine the
constraints $R_{r  z_i}=0$ which take the form
\bea \label{RrzO211} b_{11} &-& \frac{2\,\left( d-2\right)
\,c_{11}\,f + r\,a_{11}\,f'}
   {2\,\left( d-2 \right) \,f + r\,f'} - \frac{2\,\left(d-1 \right)}{d-2}
\,\frac{r\,f\,a_1\,a_1'}
   {\left[ 2\,\left( d-2 \right) \,f + r\,f' \right] }- \non
   &-& \frac{2\,r\,f\,\left( 2\,f - r\,f' \right)\,a_1\, h_1'  }{{\left[ 2\,\left( d-2\right)
\,f + r\,f' \right] }^2} -  \frac{ f\, r\,  \[2\, h_{11}' + w_{11}'+ 2\, a_{11}' +2\,(d-
2)\,c_{11}'\]}{ 2\,\left( d-
2\right) \,f + r\,f'  }=0. \eea
As usual, this equation instructs us how to choose the  gauge and
we take
\bea \label{gaugeO211T2} b_{11} &=& \frac{2\,\left( d-2\right)
\,c_{11}\,f + r\,a_{11}\,f'}
   {2\,\left( d-2 \right) \,f + r\,f'} + \non &+& \frac{2\,\left(d-1  \right)
\,r\,f\,a_1\,a_1'}
   {\left( d-2 \right) \,\left[ 2\,\left( d-2 \right) \,f + r\,f' \right]  }+
\frac{2\,r\,f\,\left( 2\,f - r\,f' \right)\,a_1\, h_1'  }{{\left[ 2\,\left( d-2\right) \,f +
r\,f' \right] }^2}.
\eea
This gauge is consistent with our basic choice (\ref{gaugeO1})
which at  this order only gets modified by adding sources from
lower orders.
 
Now (\ref{RrzO211}) reduces to $2\, h_{11}' + w_{11}'+ 2\, a_{11}'
+2\,(d-2)\,c_{11}'=0$ which we solve for  $h_{11}$
\be \label{h11} h_{11}=-\half\[ w_{11}+ 2\, a_{11}
+2\,(d-2)\,c_{11}\], \ee where the asymptotic boundary conditions
are used to eliminate the integration constant.
 
The rest of the equations are given by (\ref{O211eqs}).  Their
schematic form is
\bea \label{EqO211T2}
 {\cal E}_a (a_{11}'', a_{11}', c_{11}',  a_{11}, c_{11})&=&   Src(a_1,a_1',h_1,h_1' ) \non
{\cal E}_c (c_{11}'', c_{11}', a_{11}',  c_{11}, a_{11})&=&
Src(a_1,a_1',h_1,h_1') \non
 r^{-(d-2)} \( r^{d-2} f w_{11}'\)' &=& Src(a_1,a_1',h_1';a_{11},c_{11}),
\eea
These equations are only partially coupled and can be further
separated. The first two do not involve $w_{11}$ and moreover one
gets a decoupled equation for $a_{11}-c_{11}$. Having solved it,
we proceed to solve first for $a_{11}+c_{11}$ and then for
$w_{11}$ where at each step earlier solutions appear as sources.

The equations are subject to the horizon boundary conditions
(\ref{bcO211}) and asymptotically they must vanish. We shoot to
solve the equations, and the values of the shooting parameters are
given in table \ref{table_shootparams}. Finally, the constraint
(\ref{ConO211}) is verified to be satisfied.
 
\subsection*{Free energy and thermodynamics}
\label{sec_T2_FreeE}
In the rest of this section we change the normalization from
(\ref{diag2}) to (\ref{diag}).
 
The quadratic term in the free energy expansion remains unchanged
relative to the $\IT^1$ case: $\cA_{p=2}=\bar{\cA}=\cA_{p=1}$.

We turn to the quartic term
\be \label{F4T2}
 \bar{\cC}=F_{4GL} - \[ F_2(BR_0)+ F_2(BR_2) +F_2(BR_{11}) \]~.
\ee
We can figure out which {\it additional} terms appear in the zero
mode contribution for $\IT^2$ relative  to $\IT^1$
\be \label{F2_0Tp}
 F_2(BR_0)|_{p=2} =  F_2(BR_0)|_{p=1} - \half \int_1^{\infty} r^{d-2} \,  {h_0'}^2 dr.
\ee
The second harmonic contribution requires only a $p$-dependent
factor (division by $p=2$)
\be \label{F2_2Tp}
 F_2(BR_2)|_{p=2} = \half F_2(BR_2)|_{p=1}.
\ee
The mixed modes term, $F_2(BR_{11})$, is new, and likewise there is a new contribution to $\F4GL$. 
It is straightforward to  substitute the expansions
(\ref{O211Tp_1},\ref{O211Tp_2})  into the action integral
(remembering to divide by 4 due to change of  normalization). The
resulting expressions are cumbersome and not particularly
illuminating; we omit their  explicit form.

The numerical values of the various terms in (\ref{F4T2}) are
recorded in  table \ref{table_FT2}.
\begin{table}[t!]
\centering \noindent \hspace{-1.2cm}
\begin{tabular}{|c|c|c|c|c|c|c|c|c|c|c|c|}
\hline\hline
  \rm{d}    &4&5&6&7&8&9&10&11&12&13&14 \\ \hline
$\mathbf{\bar{\cC}} $&-.441& -1.27& -2.28& -3.24& -3.76& -3.30& -1.23&  3.17& 10.9 &23.2& 40.6
\\
$F_{4GL}$ &.2869& 2.403& 9.864& 28.97& 69.59& 145.9& 277.5& 489.4& 813.9& 1290& 1967  \\
$F_2(BR_0)$ &.2775& .8456& 1.130& -0.2807& -5.992& -19.99& -48.14& -98.35&  -180.7& -307.8& -
495.4 \\
$F_2(BR_{11})$ &.4023& 2.377& 9.121& 26.76& 65.24& 30.57& 59.29& 478.2& 804.0& 1287&1977 \\
\hline\hline
\end{tabular}
\caption[]{The non-trivial coefficients of the free energy
expansions for $\IT^2$. The rest of the coefficients are found
using the fact that $\cA$ is unchanged relative to $\IT^1$ case,
equation (\ref{F2_2Tp}) and  table \ref{table_FT1}. }
\label{table_FT2}
\end{table}
\begin{table}[t!]
\centering \noindent
\begin{tabular}{|c|c|c|c|c|c|c|c|c|c|c|c|}
\hline\hline
  \rm{d}    &4&5&6&7&8&9&10&11&12&13&14 \\ \hline
$\bar{\cC}_{p=2} $&-.441& -1.27& -2.28& -3.24& -3.76& -3.30& -1.23& 3.17&  10.9 &23.2& 40.6 \\
$\cC^{=}$ &-0.403& -1.39& -2.99& -5.03& -7.07& -8.47& -8.30& -5.27& 1.97& 15.0& 35.7 \\
$\cC^{\neq}$ &-.481& -1.14& -1.57& -1.44& -.453& 1.87& 5.83& 11.6& 19.9&  31.3& 45.5\\
\hline\hline
\end{tabular}
\caption[]{Components of the tensor $\cC$ (\ref{c-tensor}),
 that determines the order of the phase transition in the canonical ensemble. } \label{table-c}
\end{table}
\begin{table}[t!]
\centering \noindent
\begin{tabular}{|c|c|c|c|c|c|c|c|c|c|c|c|}
\hline\hline
  \rm{d}    &4&5&6&7&8&9&10&11&12&13&14 \\ \hline
$\bar{\eta}_{1,p=2} $ & 1.49& 2.81& 3.89& 4.65& 5.02& 4.84& 3.99& 2.40& -0.125& -3.73& -8.42 \\
$\eta_1^{=}$   &1.40& 3.00& 4.68& 6.21& 7.36& 7.93& 7.68& 6.29& 3.56& -0.695& -6.76  \\
$\eta_1^{\neq}$& 1.58& 2.62& 3.09& 3.10& 2.68& 1.76& 0.299& -1.48& -3.81& -6.77& -10.1\\
\hline\hline
\end{tabular}
\caption[]{Components of the tensor $\eta_1$ (\ref{eta}),
 that defines the dimensionless mass variation (see definition at the end of section
\ref{precalc-section})
 which determines the order of the phase transition in the micro-canonical ensemble.}
\label{table-eta}
\end{table}

\subsection*{$p$-dependence}
 
Table \ref{table-c} summarizes our results for the $\cC$ tensor.
For the various values of $d$, the number of extended dimensions,
we give $\cC \equiv \cC^=$ which is the $p=1$ value, together with
$\bar{\cC}|_{p=2}$. From these we determine $\cC^{\neq}$ using
(\ref{cneq}). $\cC^{\neq}$ allows to extend the results to any $p$
by computing $\bC(p)$ from (\ref{def-cbar}), which in turn
determines the order of the transition in the canonical ensemble
through (\ref{order-cond}).
 
For the micro-canonical ensemble $\cC$ is replaced by the $\eta_1$
tensor, and table \ref{table-eta} summarizes the results, which
were obtained through the use of (\ref{eta1},\ref{eta1-bar}). The
condition for a second order transition is given in
(\ref{micro-can-order-cond}).

\section{Discussion of implications}
\label{implications-section}

We demonstrated that the Landau-Ginzburg inspired method to
compute the order is somewhat simpler than the previous one, and
gives the same results (tables
\ref{table_FT1}-\ref{table_sigma-eta}). In addition we analyzed
the order of the Gregory-Laflamme transition for (square) torii
compactifications. The main results are summarized in tables
\ref{table_FT2}-\ref{table-eta}  and we now proceed to discuss
their implications.
 
\sbsection{Transition order is independent of $p$.}
A necessary condition for a second order phase transition is $0 <
\cC^= \equiv \cC$ (\ref{order-cond}). Therefore, for all $D <
D^*_{can} \equiv 12.5$, where the transition is first order for
$p=1$ (namely $\cC<0$) it is first order for all $p$.
 
From table \ref{table-c} we find that the converse it true as
well. For $d > D^*_{can}-1=11.5$ where $\cC>0$ then also
$\cC^{\neq}>0$ and therefore from (\ref{def-cbar}) also $\bC(p)
> 0$ for all $p$, and hence the transition is second order in this
range of $d$ for all $p$.
 
 Altogether we conclude that \emph{the transition order depends
only on $d$, the number of extended dimensions}, and not on $p$,
the dimension of $\IT^p$. It would be interesting to know whether
this property holds for a general compactification manifold,
$Y^p$.
 
\sbsection{Failure of the ``equal-entropy for the equal-mass
estimators''.}
Our original motivation for exploring the phase transition  order
in torus  compactification arose after observing in
\cite{KolSorkin} that the ``equal entropy for  equal mass"
estimator indicates lower critical dimension for some
higher-dimensional torii (lower then the $p=1$ value of $D^*=``13.5"$).
In fact, this argument gave an encouraging value of ``10" for
$\IT^p$ with $3 \leq p \leq 6$. However, as just noted the
critical dimension cannot be lowered, and here we
explain why the estimator failed.

We shall now show that the estimator is strongly sensitive to the
approximation error which is inherently built into it. In fact an
approximation error in the percent and sub-percent level will even
cancel the prediction of a critical dimension altogether. In other
words, it turns out that the estimates of \cite{KolSorkin} carry
unusually large error bars, which resolve the puzzle.
 
The estimator is constructed by comparing entropies of a uniform
black-brane and of a localized black-hole with the same mass. The
idea is that if entropy equality is achieved below the critical
mass this may indicate that the phase transition between these
objects is second order since a new non-uniform branch emerging
from the GL point is likely to exist and to connect between the
branches. The dimension (not necessarily integer) for which the
``equal-entropy mass'' coincides with the instability mass
estimates the critical dimension. This is only an estimate because
the entropy of a localized solution is approximated in this
approach by a naive \Schw value. This reasoning worked remarkably
well in the $\IT^1$ compactification \cite{SorkinD*}, producing a
fairly precise estimate of $D^*$.

However, the \Schw formula is only an approximation to the
localized black hole entropy, hence  we use $S_{bh}(\eta)=
S_{Schw}(\eta)(1+\eps)$, to model the  modified expression.
Perturbation theory \cite{Harmark,GorbonosKol}  and numerical
results \cite{numericII,KudohWisemanPRL} indicate that the entropy
of a localized black hole is {\it higher} than that of a \Schw
black hole and hence we must take $\eps>0$.
 
Solving for $\eta$ we get
\be \label{eta_s} \eta_S(\eps)={1\over 16 \, \pi}
\left[{\Omega_{d-2}^{~D-3} \over  \Omega_{D-2}^{~d-3} }
  {(d-2)^{(D-3)( d-2)} \over  (D-2)^{(D-2 )(d-3)}}\right]^{1/p} \,  (1+\eps)^{(D-3)(d-3)/p}
\ee
To find the estimate for the critical dimension, $\tilde{D}^*$ we
compare  this with the critical mass, which is well approximated
by the formula \cite{SorkinD*},  $\eta_c(D,p)  =\eta_c(D-p) = 0.47
\cdot 0.686^{D-p+1}$. Exploring the resulting dependence of this
estimate on $\eps$ shows that  $\tilde{D}^*$ increases fast as it
grows. For example, in a  $p=1$ case, while for $\eps=0$ one has
$\tilde{D}^*\simeq 12.5$, taking $\eps=0.002$ results in
$\tilde{D}^*\simeq 16$, and  $\eps>0.003$ does not have a
critical dimension at all. The situation is similar for other
$p$'s and for $p=6$,  $\eps \simeq 0.08$ sufficies\footnote{We see
that $\eps$ grows as $p$ increases. This can be attributed to the
fact \cite{DanPrivat} that for a given $d$ the corrections to
\Schw entropy is larger for larger $p$'s. We thank Dan Gorbonos
for sharing his unpublished results regarding small mass
corrections for caged black holes in $\IT^p$ compactifications.}
  to send $ \tilde{D}^*
\rightarrow  \infty$.
 
In conclusion, there is no contradiction between our present
results  and the estimate due to the sensitivity of the latter to
the actual value of the entropy of the localized phase.

\sbsection{Subtle spontaneous symmetry breaking.}
Examination of table \ref{table-c} reveals that for
all\footnote{The case $d=4$ is marginal. The numerical values of
$\bC|_{p=2},\cC$  are close enough, such that if both of them had
a $5\%$ numerical error (see discussion in section
\ref{sec_T2_FreeE})
the inequality would hold for $d=4$ as well.}
  $d>4$  $\bC|_{p=2} > \cC$ therefore the diagonal direction in
tachyon space is disfavored relative to turning on a ``single
tachyon'' namely a tachyon which depends only on a single $z^i$.
We interpret that to mean that the discrete symmetries of the
torus are spontaneously broken: for a first order decay the time
evolution will proceed (at least initially) through a single
tachyon, while for a second order transition the system will
re-settle into a slightly non-uniform string where mostly a single
tachyon is turned on. This makes sense if we recall that the
square torus is a special torus, and for all nearly square torii
the degeneracy of the GL modes is removed anyway (see section
\ref{precalc-section}).

\sbsection{Micro-canonical ensemble.} From table \ref{table-eta}
we find that the behavior in the micro-canonical ensemble is
similar to the case of the canonical ensemble (except for the
known change in the critical dimension). Since $\eta_1^{\neq} <
\eta_1^=$ for all $d>4$ in our table, then  $\eta_1^=$ alone
determines the order of the phase transition, and the order is
independent of $p$ (for all $d$). Moreover, as discussed in the
previous paragraph we conclude that a single tachyon is preferred
over the diagonal direction through spontaneous symmetry breaking
(for all $d>4$).

\vspace{0.5cm} \noindent {\bf Acknowledgements}
 
This research is supported in part by The Israel Science
Foundation (grants no 228/02, 607/05) and by the Binational
Science Foundation BSF-2002160, BSF-2004117. ES is supported in part by 
the Canadian Institute for Advanced Research Cosmology and Gravity Program and by 
the Natural Sciences and Engineering Research  Council of Canada.
 
The authors are grateful to the KITP in Santa-Barbara where some
final stages of this work took place. This research was supported
in part by the National Science Foundation under Grant No. PHY99-07949.

\appendix
 
\section{Second order perturbation equations}
\label{appendix_eqs}
Here we give the full back-reaction equations for each harmonic,
including the full expressions for the sources.

\sbsection{\underline{Zero-modes}}
 
First we solve the equation for $h_0$
\bea \label{Eqh0} &&{1\over r^{d-2}} \(r^{d-2} f h_0' \)'= \(
\frac{k^2\,h_1}{2\,f} +
     \frac{2\,\( d -1 \) \,\( d-3 \)\,f' \,a_1' }{{\[ 2\,\( d-2 \) \,f + r\,f'
           \] }^2} \) \frac{2\,f-r\,f'}{2\,\( d-2 \) \,f + r\,f'}\,a_1 -\non
      && \frac{4(d-1)^2(d-3)^2  \,f'}{{\[ 2\,\( d-2 \) \,f + r\,f'\] }^4}\,a_1^2 -
      \frac{k^2\[4(d-2)\,f\,\[(d^2-3d+1)\,f +d\,r\,f'\]+r^2 f'^2
\]}{2\,(d-2)\,f {\[ 2\,\( d-2 \) \,f + r\,f'\] }^2}\,a_1^2. \eea
Then the equations for $a_0$ and $c_0$ are solved
\bea
\label{Eqac0}
&&  a_0''+\frac{ 2\,{\( d-2 \) }^2\,f^2 +
       4\,\( d-2 \) \,r\,f\,f' + r^2\,f'^2  }{r\,f\, \[ 2\,\( d-2 \) \,f + r\,f'
\] }\,a_0'+\non &&
   \frac{\( d-2 \) \,f'\,
     \[ 2\,\( d-3 \) \,f + r\,f' \] }{2\,f\,
     \[ 2\,\( d-2 \) \,f + r\,f' \] }\,c_0'  +
         \frac{\( d-2 \) \,
     \( d-3 \) \,f'^2 }{f\,
     {\[ 2\,\( d-2 \) \,f + r\,f' \] }^2}\,(a_0-c_0) =
   Src_{a_0},\non
&&  c_0''+\frac{ 2\,(d-2)(2\,d-5) \,f^2 +
       4\,(d-2) \,r\,f\,f' + r^2\,f'^2  }{r\,f\,
     \[ 2\,\( d-2 \) \,f + r\,f' \] }\,c_0' +\frac{2\,(d-2)}
   {r\,\[ 2\,\( d-2 \) \,f + r\,f' \] }\,a_0'- \non &&
  \frac{2\,\,f'\,\[ (d-2)(d-3) \,{f}^2 +
       \(2\,d-5 \) \,r\,f\,f' + r^2\,{f'}^2 \] }{r\,f\,
     {\[ 2\,\( d-2 \) \,f + r\,f' \] }^2}(a_0-c_0) = Src_{c_0},
  \eea
where the sources are given by
\bea
\label{Srca0}
Src_{a_0}&=& \frac{k^2\,\( r\,f'-2\,f \) \,a_1^2}
   {2\,f\,\[ 2\,\( d-2 \) \,f + r\,f' \] }-\[ \frac{k^2\,h1}{2\,f} +
     \frac{2\,\( d -1 \) \,
       \( d-3 \)\,f' \,{a_1}' }{{\[ 2\,\( d-2 \) \,f + r\,f'
           \] }^2} \]a_1  - \frac{f'\,h_0'}{2\,f}, \\
Src_{c_0}&=&\[ \frac{k^2\,h_1}{2\,(d-2)\,f} +
     \frac{2\,(d-1)\,(d-3) \,f'\,{a_1}' }{\( d-2 \) \,
        {\[ 2\,\( d-2 \) \,f + r\,f' \] }^2} \]a_1  - \frac{h_0'}{r}+ \non
&&\frac{2\,(d-1)^2\,(d-3)\,f'^2 + (d-2)\,(2\,f-r\,f')\,\[2\,\( d-2
\) \,f + r\,f' \]\,k^2}{2\,{\( d-2 \) }^2\,f\,
     {\[ 2\,\( d-2 \) \,f + r\,f' \] }^2}  \,a_1^2
\eea
The above equations are subject to regularity boundary conditions at
the horizon
\bea \label{bcO20}
 h_0'&=&- {k^2\, a_1\, (a_1+(d-2)\, h_1) \over
2\,(d-2)\,(d-3)}, \non
 a_0'&=&-2\,(d-2)\, (a_0-c_0) + {3 \, k^2 \,a_1\,(a_1-h_1) \over 4\,(d-3) } - {(d-1)^2\,
  a_1^2 +(d-2)\,  h_0' \over 2\,(d-2)}, \non
 c_0'&=&2 \,(a_0-c_0)  - { k^2 \,a_1\,(a_1-h_1) \over 2\,(d-2)\,(d-3) }+
{(d-1)^2 \,a_1^2 \over (d-2)^2}, \eea
where both the functions and their derivatives are evaluated at $r=r_0$.
In addition, there is a constraint
 \bea \label{ConO20} &&
2\,\left( d-2 \right) \,f\,{a_0}' +
  \left( d-2 \right) \,{c_0}'\,\left[ 2\,\left( d-3 \right) \,f + r\,f' \right]   - \non
&&
  -\frac{2\,\left(d-3\right) \,\left(d-2 \right) \,\left( {a_0} - {c_0} \right) \,f'}
   {2\,\left( d-2 \right) \,f + r\,f'}  +
  \left( 2\,\left(d-2\right) \,f + r\,f' \right) \,{h_0}' +\non
  &&+
   \left( \frac{\left(d-1 \right) \,k^2\,r}{2\,\left(d-2\right) } -
     \frac{\left(d-3 \right) \,{\left(d-1 \right) }^2\,r\,{f'}^2}
      {\left( d-2 \right) \,{\left( 2\,\left( d-2 \right) \,f + r\,f' \right) }^2} \right)
{{a_1}}^2 -\frac{\left(d-1 \right) \,r\,f {a_1'} ^2}{2\,\left(d-2\right) }=0.
  \eea

\sbsection{\underline{Second harmonic}}
 
The relevant equations are
\bea
\label{O22eqs}
&&   f\,a_2'' +\frac{ (d-2) \,f + r\,f'}{r}\,a_2'  +
  \[ -4\,k^2 + \frac{2\,( d-1)(d -3) \,f'^2}{{\[ 2\, \(d-2\) \,f + r\,f'\] }^2} \]a_2
=Src_{a_2}, \non
  && {1\over r^{d-2}} \(r^{d-2} f h_2' \)'  = Src_{h_2}.
\eea
The sources read
\bea
\label{SrcO22}
Src_{a_2}&=&-3\,\[ \frac{(d-1)(d-3)f f' \,a_1'}{\[ 2\,\( d-2 \) \,f + r\,f' \]^2} +\half
\,h_1\,k^2\]\,a_1-\non
&-& \[ \frac{(d-1)^2\,(d-3) \,r \,f'^3}{(d-2) \[ 2\, \(d-2\) \,f + r\,f'\]^3}+
\frac{3\,k^2\,(2\,f-r\,f')}{ 2\,\[ 2\, \(d-2\) \,f + r\,f'\]}\]\,a_1^2,\non
Src_{h_2}&=& \frac{2\,(d-1)\,(d-3)\,f \,f' (2\,f-r\,f')\,a_1'\,a_1}{\[ 2\,\( d-2 \) \,f +
r\,f' \]^3} -3\,(d-2)\,k^2\, h_1 \,a_1  +\non
&+& \frac{4\,k^2\,(d-1)\,r\,f\,a_1'}{\[ 2\,\( d-2 \) \,f + r\,f' \] } \,a_1+
\frac{ 4\,(d-3)^2(d-1)^2f\,f'^2}{\[ 2\,\( d-2 \) \,f + r\,f' \]^4}\,a_1^2  + \non
&- &\[ \frac{k^2 \(4(d-2)f\[(d^2-3\,d+5)\,f+(d-4)\,r\,f'\]+(4\,d-7)r^2 f'^2\)}{\[ 2\,\( d-2 \)
\,f + r\,f' \]^2} \]\,a_1^2  + \non
&+& 4\,k^2\,\frac{2\,f - r\,f' }{2\,\(d-2\) \,f + r\,f'}\,a_2
\eea
These equations are subject to regularity boundary conditions at the horizon
\bea
\label{bcO22}
a_2'&=& -2\,\(  d-1 - {2\,k^2\over d-3} \) a_2 + \[ {(d-1)^2 \over
  d-2}- {3\,k^2 \over 2\,(d-3)}\]\,a_1^2 +{3\,k^2 \,a_1\,h_1 \over
  2\,(d-3)}, \non
h_2'&=&  { 8\,(d-2)\,a_2 +a_1\,\[ (4\,d-7) a_1 - 3\,(d-2)\,h_1\] \over
  2\,(d-2)\,(d-3)}\, k^2,
\eea
where both the functions and their derivatives are computed at
$r=r_0$.
 
The constraint is
\bea
\label{ConO22}
&&h_2'+ \frac{a_2'\,\left( 2\,f - r\,f' \right) }{2\,\left( d-2 \right) \,f + r\,f'}+
   \frac{2\,\left( d-3 \right) \,\left( d-1\right) \,f'\,{a_2}}
   {{\left( 2\,\left( d-2 \right) \,f + r\,f' \right) }^2} +\non
&&-\frac{\left( d-1 \right) \,\left(d-3 \right) \,f\,a_1\,a_1'}
   {{\left( 2\,\left( d-2 \right) \,f + r\,f' \right) }^2}  -
    \frac{\left( d-1 \right) \,r\,f\,{a_1'}^2}
   {2\,\left( d-2 \right) \,\left( 2\,\left( d-2 \right) \,f + r\,f' \right) }  - \non
   &&-\left[ \frac{\left(d-3 \right) \,{\left( d-1\right) }^2\,r\,f'^2}
      {\left( d-2 \right) \,{\left( 2\,\left( d-2 \right) \,f + r\,f' \right) }^3} +
     \frac{\left( d-1\right) \,k^2\,r}{2\,\left( d-2 \right) \,\left( 2\,\left( d-2 \right) \,f
 + r\,f' \right) }
     \right] \,a_1^2 =0.
     \eea
     %
 
\sbsection{\underline{Mixed modes}}
 
The mixed modes back-reaction appears in the $\mathbf{T}^p$
calculation starting from $p=2$. The perturbations are governed by
the equations
\bea
\label{O211eqs}
f\, a_{11}''&+&{(d-2) f + r\, f' \over r} a_{11}' - 2\,k^2\,a_{11} +
{2(d-2)(d-3)f'^2\over \[2\,(d-2)\,f +r\,f'\]^2}\,(a_{11}-c_{11})
+Src_{a_{11}}=0,\non
f\, c_{11}''&+&{(d-2) f + r\, f' \over r} c_{11}' - 2\,k^2\,c_{11} -
{2(d-3)f'^2\over \[ 2\,(d-2)\,f+r\,f' \]^2 }\,(a_{11}-c_{11})
+Src_{c_{11}}=0, \non
f\, w_{11}''&+&{(d-2) f + r\, f' \over r} w_{11}'+Src_{w_{11}}=0.
\eea
where
\bea
\label{SrcO211}
Src_{a_{11}}&=&\frac{4 (d-1)^2 f'^2 \left(-2 (d-3) (d-2) f^2-(d-1) r
   f' f+r^2 f'^2\right) }{(d-2) \left(2
   (d-2) f+r f'\right)^3}\,a_1^2+\non
&+&\frac{k^2 \left(4 (d-2) f
   \left(8 (d-2)^2 f^2-\left(5 d^2-27 d+33\right) r f'
   f-5 (d-2) r^2 f'^2\right) \right)}{(d-2) \left(2 (d-2) f+r
   f'\right)^3}\,a_1^2-\non
&-&\frac{(4 d-7) r^3 f'^3
   }{(d-2) \left(2 (d-2) f+r
   f'\right)^3}\,a_1^2+\frac{4 (d-1) f f'}{2 (d-2) f+r f'}\,a_1'a_1 +\frac{2
   f  \left(2 f-r f'\right)}{2 (d-2)
   f+r f'}a_1'^2+\non
&+&\frac{2 f \left(4 (d-2) \left(2
   d^2-5 d+1\right) f^2+\left(5 d^2-6 d-7\right) r f'
   f-(d-3) r^2 f'^2\right) f' }{\left(2
   (d-2) f+r f'\right)^3}\,h_1'a_1-\non
&-&\frac{r f f' }{2 (d-2)
   f+r f'}\,h_1'^2+\frac{f  \left(3 r^2 f'^2+2
   f \left(4 f (d-2)^2+(4 d-9) r f'\right)\right)
  }{\left(2 (d-2) f+r f'\right)^2} \,a_1'h_1', \non
Src_{c_{11}}&=&\frac{4 (d-1)^2 f'^2 \left(2 (d-3) (d-2) f^2+(d-1) r
   f' f-r^2 f'^2\right) }{(d-2)^2
   \left(2 (d-2) f+r f'\right)^3}\,a_1^2+\non
&+&\frac{2 k^2 \left(2
   (d-2) r^3 f'^3+(2-d) \left(4 (d-2) f^2
   \left(\left(d^2+d-7\right) f+(8-d) r f'\right)\right)\right)}{(d-2)^2
   \left(2 (d-2) f+r f'\right)^3}\,a_1^2-\non
&-&\frac{(8
   d-21) r^2 f f'^2 }{(d-2)^2
   \left(2 (d-2) f+r f'\right)^3}\,a_1^2-\frac{4 (d-1) f
   f'}{(d-2) \left(2 (d-2) f+r
   f'\right)}\,a_1' a_1-  \non
&-&\frac{4 f\left((d-2) \left(5 d^2-14 d+5\right) f^2+4 (d-2) d r
   f' f+r^2 f'^2\right) f' }{(d-2)
   \left(2 (d-2) f+r f'\right)^3}\,h_1' \,a_1-\non
&-&\frac{2 f^2
   }{2 (d-2) f+r f'}\,h_1'^2-\frac{2 f
  \left(r^2 f'^2+(d-2) f \left(2 (2 d-3)
   f+3 r f'\right)\right) }{(d-2) \left(2
   (d-2) f+r f'\right)^2}\, a_1' h_1'+\non
&+&\frac{2 f
   \left(r f'-2 f\right)}{(d-2) \left(2 (d-2) f+r
   f'\right)}\,a_1'^2,\non
Src_{w_{11}}&=&2 (d-2) {c_{11}} \left(1+\frac{2 f}{2 (d-2) f+r
   f'}\right) k^2+4 {a_{11}} \left(1-\frac{(d-2)
   f}{2 (d-2) f+r f'}\right) k^2+\non
&+&\frac{2
   \left[(2\, d-3) r^2 \,f'^2+4\, (d-2)\, f
   \left(\left[d^2-3\, d+3\right]\, f+(d-2)\, r
   f'\right)\right] k^2}{(d-2) \left(2 \,(d-2)\, f+r
   f'\right)^2}\,a_1^2 +\non
&+& 4  \left[\frac{(d-1) \,r \,f
   }{(d-2) \left(2 \,(d-2) \,f+r
   f'\right)}\,a_1'\,a_1-\frac{r\, f \,\left(r \,f'-2 \,f\right)
   }{\left[2 \,(d-2) \,f+r \,f'\right]^2}\,h_1'\,a_1\right]
   k^2 .
\eea
 
The horizon boundary conditions for these equations are given by
\bea
\label{bcO211}
a_{11}'&=&\frac{\left[(4 d-7) k^2-4 (d-3) (d-1)^2\right]
   a_1^2+2 (d-2) \left[\left(k^2-(d-5) d-6\right)
   a_{11}\right]}{(d-3) (d-2)} +c_{11},\non
c_{11}'&=&\frac{2}{d-3}\, \left(\frac{2 \left[(d-3) (d-1)^2-(d-2)
   k^2\right] a_1^2}{(d-2)^2}+\left(k^2-d+3\right)
   c_{11}\right) +a_{11},\non
w_{11}'&=&-\frac{2 \,k^2 \left[(2\, d-3)\, a_1^2+(d-2)\,
   (2\, a_{11}+(d-2)\, c_{11})\right]}{(d-3) (d-2)}.
\eea
where the functions and the derivatives are evaluated at $r=r_0$.
 
Finally, the constraint for the mixed modes reads
\bea
\label{ConO211}
&&w_{11}'+
\frac{2  \left[(d-2)  f+r f'\right]a_{11}'}{2 (d-2) f+r f'}+
\frac{(d-2) \left[2 (d-1) f+r f'\right]c_{11}'}{2 (d-2) f+r f'}
+\frac{2 (d-3) (d-2)f' ({a_{11}}-{c_{11}})}{\left(2 (d-2) f+r  f'\right)^2}+\non
&&\frac{2 (d-1) r f {a_1'}^2}{(d-2) \left(2 (d-2) f+r f'\right)}
+ \frac{4 (d-3) (d-1) f a_1 a_1'}{\left(2\,(d-2) f+r f'\right)^2}
-\frac{2 r f {h_1'}^2}{2 (d-2) f+r f'} -\non
&&\frac{4 \,(d-2)  f \left(-2 (d-3) f^2+(d-5) r f' f+r^2 f'^2\right)
  a_1 h_1'}{\left(2 (d-2) f+r f'\right)^3} +
\frac{4 (d-3) (d-1)^2 r f'^2 a_1^2}{(d-2) \left(2 (d-2) f+r f'\right)^3}=0.
\eea
%
\section{Numerical issues and shooting parameters}
\label{appendix_numerics}
In this appendix we give some details regarding our numerical
implementation and describe several tests that demonstrate its
robustness. In addition, we summarize the numerical values of the
shooting parameters in table \ref{table_shootparams}.
 
The ODEs encountered in this paper must be solved between the
horizon and infinity, $ r \in [r_0,\infty)$. However, in practice
we solve the equations only  for $ r \in [r_0+\eps,r_{max}]$,
where $\eps \ll r_0$ and $r_{max}$ is some finite number. The
first boundary is taken slightly outside $r_0$ because the
equations become singular at the horizon. To get the correct
boundary conditions at this point we Taylor-expand our functions
to the third order about the horizon and find the expansion
coefficients analytically by plugging the expansion into the
equations. This enables us to compute the functions and their
derivatives at $r_{min}=r_0+\eps$. We checked that the meaningful
numbers cited in this paper are independent  of $\eps$ (with
better then $0.1\%$ precision) provided $\eps \leq 10^{-4}$. As
for the outer boundary,  its location is determined by  the
``empirical formula'', $r_{max} (d) \simeq 18-d$ in the range
$4\leq d \leq 14$. Here too the meaningful numbers undergo less
then $0.1\%$ variations when we tried larger $r_{max}$.
 
In our shooting routines we integrate the equations from
$r_{min}$, where one of the boundary conditions (b.c.) is used,
toward $r_{max}$ where we attempt to satisfy the asymptotic b.c.
For $k \neq 0$ this
integration will in general diverge when $ r_{max} \rightarrow
\infty$ since both the decaying and the growing solutions are
present\footnote{For a finite $r_{max}$ the result of this
integration is a finite number that grows roughly exponentially
with $r_{max}$.}. Only for a specific value of the shooting
parameter the growing solution is eliminated, the integration
converges and asymptotic b.c. is satisfied. Our method to filter
out this growing solution follows that of Gubser \cite{Gubser}: we
use the integral of  $|x(r)|^2$ near the outer boundary, where
$x(r)$ is the mode we are solving for, as an indicator for how
well the asymptotic b.c. is satisfied (for the true solution this
integral must be negligible). The shooting and matching procedure
is thus reduces to a minimization of that integral.
 
The situation is different for the zero modes, as for them a
finite solution exists regardless of the value of the shooting
parameters. In this case we demand  that the solution vanishes
asymptotically. This is easily accomplished for $h_0$, for whom
``the second shot hits". However, as described in section
\ref{sec_T1_O2}, the procedure should be iterated to make $a_0$
and $c_0$ vanish asymptotically. Moreover, because of the
relatively slow power-law decay the location of the  outer
boundary at $r_{max}$, given by the above empirical formula, is
insufficient to ensure this asymptotic vanishing accurately
enough. So, in practice, after integrating the equations
(\ref{Eqac0}) to $r_{max}$ we set the sources to zero (since they
decay exponentially and are already very small) and continue to
integrate to some large $r_{asymp}$ (taken to be an order of
magnitude larger than $r_{max}$.) In the vicinity of $r_{asymp}$
we fit the solutions by expressions of the form $a_\infty
+\alpha/r^{d-2}$ and $c_\infty +\gamma/r^{d-2}$, which is the
correct analytic behavior in the asymptotic region
\cite{numericI}. Then we compute $\Delta = |c_\infty|+|a_\infty|$,
which is an effective measure for how well the boundary conditions
are satisfied. To minimize $\Delta$ we iterate until $\Delta$
decreases below a certain tolerance.

The numerical values of all shooting parameters appearing in this
paper are listed in table \ref{table_shootparams}.

As soon as the solutions are obtained one must ensure that the
constraints (\ref{ConO1}), (\ref{ConO20}), (\ref{ConO22}) and
(\ref{ConO211}) are satisfied by these solutions. In our case we
verified that the constraints are satisfied with better then
$0.1\%$ precision. This together with other tests described above
gives an idea of the numerical accuracy of our method.

\begin{table}[t!]
\centering
\noindent
\begin{tabular}{|c|c|c|c|c|c|c|c|c|c|c|c|}
\hline\hline
  \rm{d}  &4&5&6&7&8&9&10&11&12&13&14 \\  \hline\hline
$h_1(1)$ &.434&.496&.532& .560& .576& .595& .604& .613& .620& .627& .631 \\
$h_0(1)$ &.796& 1.01& 1.24& 1.48& 1.73& 1.96& 2.21& 2.46& 2.71& 2.95& 3.21  \\
$a_0(1)$ &-.232& .015& .506& 1.24& 2.23& 3.46& 4.95& 6.68& 8.68& 10.9& 13.4 \\
$c_0(1)$ &-.730& -.921& -1.14& -1.38& -1.62& -1.86& -2.11& -2.36& -2.60& -2.85& -3.10 \\
$a_2(1)$ &-.40& .690& 1.11& 1.66& 2.37& 3.22& 4.22& 5.43& 6.76& 8.26& 9.93  \\
$h_2(1)$ &-.812& -1.22& -1.65& -2.09& -2.54& -2.99& -3.46& -3.89& -4.38& -4.84&-5.47 \\
$a_{11}(1)$ &5.28& 7.74& 10.2& 12.5& 14.8& 17.1& 19.8& 22.2& 24.7& 27.1& 29.6 \\
$c_{11}(1)$ &-2.96& -2.92& -2.90& -2.88& -2.86& -2.84& -2.88& -2.89& -2.89& -2.90& -2.90 \\
$w_{11}(1)$ &2.96& 4.01& 5.04& 6.04& 7.03& 8.01& 9.08& 10.1& 11.0& 12.1& 13.1 \\
\hline\hline
\end{tabular}
\caption[]{The shooting parameters in various dimensions. We
assume here that the normalization for any $p$ is same as for
$\IT^1$.  The last 3 lines appear
only for torii ${\bf T}^p$ with  $p \geq 2$. }
\label{table_shootparams}
\end{table}
%
 

\end{document}